\documentclass[showpacs,twocolumn,aps,nofootinbib]{revtex4-1}
\usepackage[english]{babel}
\usepackage[utf8]{inputenc}
\usepackage{graphicx}
\usepackage{url}
\usepackage[thinlines]{easytable}
\usepackage{epstopdf}
\usepackage{amsmath, amssymb}
\usepackage{dcolumn}
\usepackage[mathscr]{euscript}
\usepackage{caption}
\usepackage{subcaption} 
\usepackage{amsmath,amsfonts,amsthm,bm}
\usepackage{mathtools}
\usepackage{float}
\usepackage{color}
\usepackage{braket}
\usepackage{wasysym}

\DeclareMathOperator*{\SumInt}{%
	\mathchoice%
	{\ooalign{$\displaystyle\sum$\cr\hidewidth$\displaystyle\int$\hidewidth\cr}}
	{\ooalign{\raisebox{.14\height}{\scalebox{.7}{$\textstyle\sum$}}\cr\hidewidth$\textstyle\int$\hidewidth\cr}}
	{\ooalign{\raisebox{.2\height}{\scalebox{.6}{$\scriptstyle\sum$}}\cr$\scriptstyle\int$\cr}}
	{\ooalign{\raisebox{.2\height}{\scalebox{.6}{$\scriptstyle\sum$}}\cr$\scriptstyle\int$\cr}}
}

\newcommand{\be}{\begin{equation}}
\newcommand{\ee}{\end{equation}}
\newcommand{\bq}{\begin{eqnarray}}
\newcommand{\eq}{\end{eqnarray}}

\begin{document}
\renewcommand{\figurename}{FIG.}

\title{Relativistic Effects in Spin Correlations Induced by QED Scattering and Wigner Rotations}

\author{Juan D. Fonseca$^a$}\email[]{juan.fonseca@ufabc.edu.br}
\author{B. Hiller$^b$}\email[]{brigitte@fis.uc.pt}
\author{I. G. da Paz$^c$}\email[]{irismar@ufpi.edu.br}
\author{M. Sampaio $^a$}\email[]{marcos.sampaio@ufabc.edu.br}

\affiliation{$^{a}$  CCNH, Universidade Federal do ABC,  09210-580 , Santo Andr\'e - SP, Brazil}
\affiliation{$^{b}$  CFisUC - Department of Physics, University of Coimbra, 3004-516 Coimbra, Portugal}
\affiliation{$^c$ Universidade Federal do Piau\'{\i}, Departamento de F\'{\i}sica, 64049-550, Teresina, PI, Brazil}

\begin{abstract}
\noindent
We study the relativistic nature of the interactions that, at tree level, generate spin correlations between two electrons in Møller scattering, as well as in an extended process involving a witness particle $C$. The corresponding processes, $e^{-}e^{-}\rightarrow e^{-}e^{-}$ and $e^{-}e^{-}C\rightarrow e^{-}e^{-}C$, are analyzed both in the center-of-mass frame and, for the former process, in a Lorentz-boosted frame where Wigner rotations arise. It is found that, through a nonrelativistic approximation of the scattering amplitudes, dipole–dipole and current–dipole interactions are responsible for the emergence of these correlations. This is evidenced by the variation of the von Neumann entropy of one electron for initially separable
states, and of $C$ for an initially prepared three-particle entangled W-state. In Wigner rotations, the invariance of entropy under local unitary transformations is maintained at the expense of the emergence of quantum coherence in the density matrix at large rapidities. As a consequence, the final states of both particles are evaluated and shown to encode information
about the scattering process through their spin expectation values. This framework is then used to comment on the correlations in the inelastic process $e^{-}e^{+}\rightarrow\mu^{-}\mu^{+}$, for which some research has reported differing results.

\end{abstract}

\maketitle

\section{Introduction}

Understanding how quantum information behaves in relativistic settings is essential for uncovering the fundamental structure of physical reality. While non‑relativistic quantum mechanics successfully captures many low‑energy phenomena, it becomes inadequate in regimes where high energies, strong gravitational fields, or accelerated frames play a decisive role. In such contexts relativistic effects can profoundly modify both the generation and degradation of entanglement.
A central example is the Unruh effect \cite{unruh}, in which uniformly accelerated observers perceive the inertial vacuum as a thermal state. This phenomenon alters the correlations shared between field modes and consequently reshapes the entanglement structure observed by different frames. Similarly, relativistic scattering processes offer a natural arena for studying how entanglement is produced, redistributed, or destroyed at high energies. Investigations of entanglement in particle collisions  provide insight into our understanding of quantum field theory, but also highlight the role of entanglement as a physically meaningful and experimentally relevant quantity in relativistic physics. 

Within the study of entanglement production and/or annihilation in scattering processes, multiple scenarios  are considered, such as the underlying interactions, relativistic effects, and the inclusion of additional particles that do not directly participate in the process. For example, the increase in the entanglement entropy (EE) in a collision between a pair of fermions when their initial state does not have a well-defined helicity is computed in \cite{serafini}. Ref. \cite{Feder} examines the entanglement and entropy of a process with spin-dependent and spin-independent states, introducing impact parameters under a Coulomb interaction. Ref. \cite{Seki} analyzes the EE using a partial-wave expansion over an initial state defined in momentum space. Unlike \cite{Seki}, Ref. \cite{Seki 2} computes the variation of momentum entropy for both a separable initial state and an entangled one. The variation of EE using an initial entangled spin state parameterized by two variables is evaluated in \cite{jimbo}.

On the other hand, the implications of collisions between indistinguishable particles for entanglement generation have also motivated the consideration of additional effects at the relativistic frontier. For instance, Refs. \cite{Lamata} and \cite{Kouzakov} study the entanglement between two identical fermions induced by the antisymmetrization of their states, generated by a spin-independent interaction (within a non-relativistic regime in the latter case). Ref. \cite{Pachos} addresses the entanglement generated by a spin-dependent interaction, extending the analysis to a relativistic framework through a spinorial representation.

Other works further develop these relativistic aspects of entanglement production and/or annihilation by exploiting the fact that spin labels the irreducible representations of the little group associated with a fixed four-momentum in the Poincaré group. An especially interesting effect is given by Wigner rotations, induced by an observer moving perpendicular to the particle momenta $\bf{p}$, as measured in a rest frame, who perceives a rotation in the spin parameterized by the momentum in the new reference frame $\Lambda\bf{p}$. For example, the entanglement that arises between different partitions of a two-particle system when a perpendicular boost is applied to their motion is analyzed in \cite{Bertlman}. In addition, Ref. \cite{Alsing} analyzes how spinors transform under Wigner rotations, while Ref. \cite{Jimbo2} investigates the resulting entanglement structure when such rotations act on both particle states and scattering amplitudes. The latter reference examines the entanglement under boosts parallel to the initial direction of motion.

Finally, other works examine the correlations generated in scattering processes involving a spectator particle. For example, Refs. \cite{Jonas,Fonseca} derive the correlations developed by that particle at tree level and their proportionality to the cross section. Refs. \cite{Blasone, blasone2, Shanmuka} study the distribution of entropy quantum information among different subsystems that include the spectator particle, with unitarity of the scattering operator $\mathcal{S}$ imposed in the latter reference.

In this work, we investigate the relativistic nature of the interactions that give rise to correlations in a scattering process involving both interacting and spectator particles, as well as the conditions that initial states must satisfy for such correlations to emerge. We further analyze the role of Wigner rotations in the creation of quantum coherence through correlations developed along directions transverse to the initial motion, together with the change in mixedness caused by the scattering process. To this end, the paper is organized as follows: In Sec. \ref{scatteringt}, we construct the tree-level reduced density matrix for the process $e^{-}e^{-}\rightarrow e^{-}e^{-}$ using the scattering formalism and Wigner rotations. In Sec. \ref{aproximation}, the non-relativistic approximation  of Feynman amplitudes is made through the expansion of the particle momenta to zeroth- and first- order. In Sec. \ref{2-2}, two initial states for the interacting particles are considered, one entangled and other separable, to analyze the correlations generated by the scattering amplitudes and Wigner rotations. In Sec. \ref{testemunha}, the procedure is repeated with the inclusion of a spectator particle prepared in a tripartite W state. In Sec. \ref{inelastic}, we use the last results to make some remarks on the correlations arising in the inelastic scattering $e^{-}e^{+}\rightarrow\mu^{-}\mu^{+}$. Finally, conclusions and perspectives are left to Sec. \ref{conclusions}. 

\section{scattering of two fermions at tree-level}\label{scatteringt}
\subsection{Center of mass frame}
As usual, asymptotic particle states in the distant past and future ($t\rightarrow \pm \infty$)
are described as wave packets that are well separated in the position space and narrowly peaked in the momentum space. Here, we take them to be plane waves with definite momenta. The unitary scattering operator $\widehat{\mathcal{S}}$ is defined as the time evolution in the interaction
picture in terms of asymptotic states. Therefore, the initially factorized state and its corresponding out-state are
\begin{eqnarray}
\ket{\psi_{\text{in}}} &=& \bigotimes_{j=1}^n\ket{\textbf{\textup{p}}_{\textup{i}}^j,s_{\textup{i}}^j},\nonumber\\
\ket{\psi_{\textup{out}}} &=& \mathscr{N}^{-1}\bigg[\bigotimes_{k=1}^{m}\mathcal{\widehat{I}}^{(k)}_{\textup{f}}\bigg]\widehat{ \mathcal{S}}\ket{\psi_{\textup{in}}},\label{in-out kets}
\end{eqnarray}
where $n(m)$ are the initial (final) particles, and
\be
\mathcal{\widehat{I}}_{\textup{f}} =\SumInt_{\bf{p}_{f},s_{f}}\ket{\textbf{\textup{p}}_{\textup{f}},s_{\textup{f}}}\bra{\textbf{\textup{p}}_{\textup{f}},s_{\textup{f}}}.\label{completeness}
\ee
 For short hand notation, we use $\int_{\textbf{\textup{p}}_{\textup{f}}}\equiv[(2\pi)^{3}2E_{\textbf{\textup{p}}_{\textup{f}}}]^{-1}d^{3}\textbf{\textup{p}}_{\textup{f}}$ as the Lorentz invariant measure over final momentum. The sum $\sum_{s_{\textup{f}}}$ runs over the final spin states and the factor $\mathscr{N}^{-1}$ is such that $\ket{\psi_{\textup{out}}}$ is normalized. It is useful to define the transition operator $\widehat{\mathcal{T}}$ through $\widehat{\mathcal{S}}=\widehat{\mathcal{I}}+i\widehat{\mathcal{T}}$ which accounts for the non-trivial scattering operator, and is connected to the transition amplitude $\mathcal{M}$ as determined by the Feynman diagram. The corresponding expression reads
\begin{eqnarray}
&&\braket{\textbf{\textup{p}}_{\textup{f}}^{(1)},s_{\textup{f}}^{(1)};\textbf{\textup{p}}_{\textup{f}}^{(2)},s_{\textup{f}}^{(2)}|i\widehat{\mathcal{T}}|\textbf{\textup{p}}_{\textup{i}}^{(1)},s_{\textup{i}}^{(1)};\textbf{\textup{p}}_{\textup{i}}^{(2)},s_{\textup{i}}^{(2)}}= \nonumber \\  i&&(2\pi)^{4}\delta^{4}(p_{\text{i}}^{\alpha(1)}+p_{\text{i}}^{\alpha(2)}-p_{\text{f}}^{\alpha(1)}-p_{\text{f}}^{\alpha(2)})\mathcal{M}_{\text{i}\rightarrow\text{f}},\label{ToperatorM}
\end{eqnarray}
for $n=m=2$, and
\be
\mathcal{M}_{\text{i}\rightarrow\text{f}}=\mathcal{M}^{\textbf{\text{p}}_{\text{i}}^{(1)},\textbf{\text{p}}_{\text{i}}^{(2)}}_{\textbf{\text{p}}_{\text{f}}^{(1)},\textbf{\text{p}}_{\text{f}}^{(2)}}(s_{\text{i}}^{(1)}s_{\text{i}}^{(2)}\rightarrow s_{\text{f}}^{(1)}s_{\text{f}}^{(2)}).\label{Mvectors}
\ee
This allows us to construct the total density matrix
\be
\rho_{\text{out}}= \mathcal{N}^{-1}\ket{\psi_{\text{out}}}\bra{\psi_{\text{out}}},\label{totalDensity}
\ee
and the reduced density matrix of \textit{one} particle
\be 
\rho_{one}=\text{Tr}_{other}[\rho_{\text{out}}],
\label{matricesreducidas}
\ee
 where the trace is taken over the \textit{other} particles using $\text{Tr}[\rho]=\sum_{\sigma}\int_{\textbf{\text{k}}}\bra{\textbf{\text{k}},\sigma}\rho\ket{\textbf{\text{k}},\sigma}$. 
The normalization of the total density matrix, $\mathcal{N}$, in accordance with $\text{Tr}[\rho_{\text{out}}]=1$, is calculated by tracing \textit{all} particles: $\mathcal{N}=\text{Tr}_{all}[\rho_{\text{out}}]$. Also, it is useful to write 
\be
\mathcal{N}=\mathcal{N}_{\text{in}}+\mathcal{N}_{\text{cross}}+\mathcal{N}_{\text{trans}},\label{TotalN}
\ee
with
\bq
\mathcal{N}_{\text{in}}&=&\text{Tr}_{all}(\ket{\psi_{\text{in}}}\bra{\psi_{\text{in}}}), \nonumber\\
\mathcal{N}_{\text{cross}}&=&\text{Tr}_{all}(\ket{\psi_{\text{in}}}\bra{\psi_{\text{trans}}}+\ket{\psi_{\text{trans}}}\bra{\psi_{\text{in}}}),\nonumber\\
\mathcal{N}_{\text{trans}}&=&\text{Tr}_{all}(\ket{\psi_{\text{trans}}}\bra{\psi_{\text{trans}}}).
\label{partialNormalizations}
\eq
The state $\ket{\psi_{\text{trans}}}$ 
is obtained by the action of $\widehat{\mathcal{T}}$ on the initial state and depends on the specific scattering process under consideration. 
On the other hand, unitarity of the scattering operator implies
\begin{equation}
\widehat{\mathcal{I}}+i(\widehat{\mathcal{T}}-\widehat{\mathcal{T}}^{\dagger})+\widehat{\mathcal{T}}^{\dagger}\widehat{\mathcal{T}}=\widehat{\mathcal{I}}\label{unitarity}.
\end{equation}
This relation shows that the terms $i(\widehat{\mathcal{T}}-\widehat{\mathcal{T}}^{\dagger})$ and $\widehat{\mathcal{T}}^{\dagger}\widehat{\mathcal{T}}$  must cancel order by order in perturbation theory. At tree level, the scattering amplitudes are real, so the imaginary part of the forward amplitude vanishes. However, the term $\widehat{\mathcal{T}}^{\dagger}\widehat{\mathcal{T}}\sim|\mathcal{M}|^{2}$ contributes at order $e^{4}$. Therefore, loop corrections to $\hat{\mathcal{T}}$ are required in order to restore unitarity at that order. This is precisely the content of the optical theorem. When loop corrections are included, it is also necessary to consider real photon emission processes, $A+B\rightarrow A+B+\gamma$, since infrared divergences arise and must cancel in inclusive observables such as the cross section. In what follows, we restrict ourselves to the (1+1)-particle sector of the Hilbert space $\mathcal{H}_{A}\otimes\mathcal{H}_{B}$ and assume weak coupling, so that unitarity is satisfied perturbatively at tree level,
\begin{align}
\widehat{\mathcal{S}}^{\dagger}\widehat{\mathcal{S}}=\widehat{\mathcal{I}}+\mathcal{O}(e^{4}).
\end{align}
This is important when calculating $\mathcal{N}_{\text{trans}}$, since its values are canceled by loops that arise in $\mathcal{N}_{\text{cross}}$. This implies that the total normalization of any density matrix consists only of $\mathcal{N}_{\text{in}}$. This was one of the results discussed in \cite{Shanmuka}, in which the reduced density matrix of a witness $C$ remained unchanged by scattering even when it began entangled with one of the interacting particles. However, in what follows, these higher orders in perturbation theory will be omitted, so the total normalization is given by  $\mathcal{N}_{\text{in}}+\mathcal{N}_{\text{trans}}$, and $\mathcal{N}_{\text{cross}}=0$.\\ 

\begin{figure}[H]
\captionsetup{justification=centering}
\centering
\includegraphics[scale=0.5]{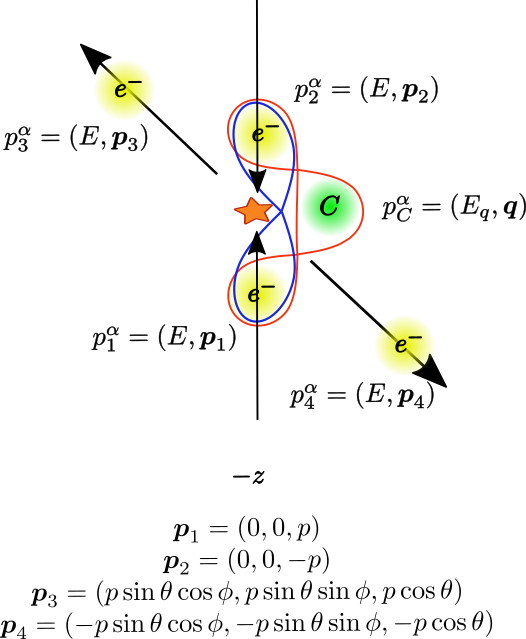} 
\caption{Collision diagram for the elastic scattering $e^{-}e^{-}\rightarrow e^{-}e^{-}$.}
\label{fig:Moller general}
\end{figure}

\begin{figure}[H]
\centering
\captionsetup{justification=centering}
\begin{subfigure}[H]{0.48\textwidth}
\centering
\includegraphics[width=6.6cm, height=5.7cm]{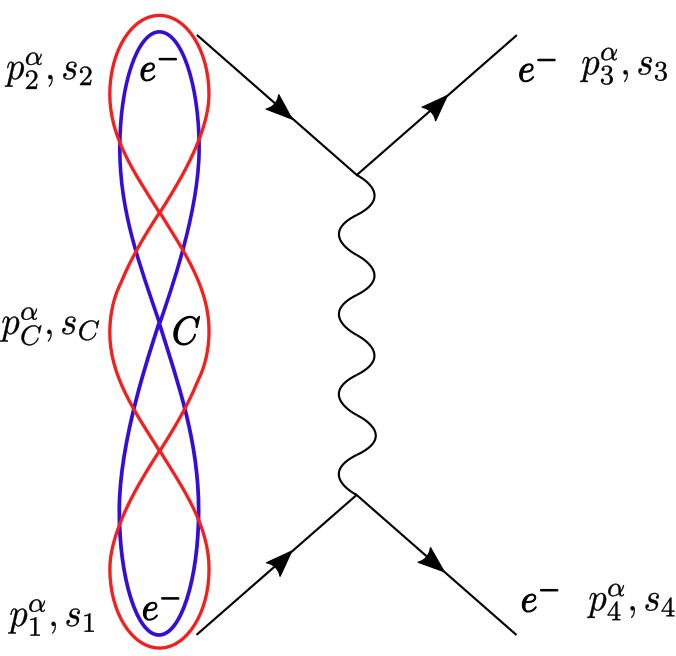}
\caption{The \textit{t} channel for M\o ller scattering $e^{-}e^{-}\rightarrow e^{-}e^{-}$.}
\label{fig:f5}
\end{subfigure}
\hfill
\begin{subfigure}[H]{0.48\textwidth}
\centering
\includegraphics[width=6.6cm, height=5.7cm]{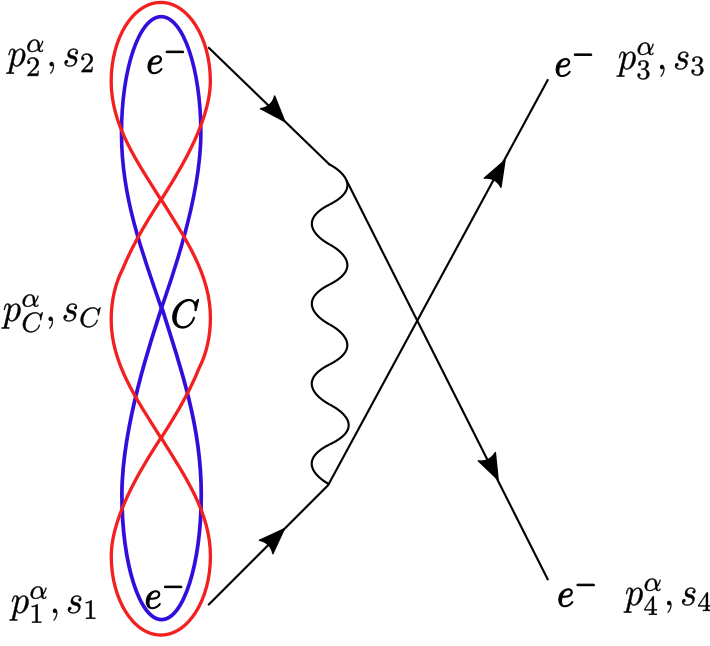}
\caption{The \textit{u} channel for M\o ller scattering $e^{-}e^{-}\rightarrow e^{-}e^{-}$.}
\label{fig:f6}
\end{subfigure}
\caption{Feynman diagrams for M\o ller scattering.}
\label{fig:feyman}
\end{figure}

Consider the M\o ller scattering in quantum electrodynamics, $e^{-}e^{-}\rightarrow e^{-}e^{-}$,  taken in the center-of-mass frame where the electron collision occurs along the $z$ axis with momentum $p$ and energy $E$, as depicted in Fig. \ref{fig:Moller general}.  
The scattering angle between $\bf{p_1}(\bf{p_2})$ and $\bf{p_3}(\bf{p_4})$ is $\theta$  and $\phi$ is the azimuthal angle. In section \ref{2-2}, we will consider two electrons which may be entangled in spin degrees of freedom, as represented by the blue curved line through the state $\ket{\Psi^{+}}=(2)^{-1/2}[\ket{\uparrow\downarrow}+\ket{\downarrow\uparrow}]$; in section \ref{testemunha}, we will admit a spectator particle $C$ (Claire)  which is entangled, also in spin degrees of freedom, with both electrons as represented by the red curved line through the state $\ket{\text{W}}=(3)^{-1/2}[\ket{\downarrow\downarrow\uparrow}+\ket{\downarrow\uparrow\downarrow}+\ket{\uparrow\downarrow\downarrow}]$  \cite{vidal}. For specific spin quantum numbers $s_i$, the tree-level contribution to this scattering contains the difference between $t$ and $u$ channels represented, respectively, by the Feynman diagrams depicted in Fig. \ref{fig:feyman}.
The analytic expression reads
\begin{align}
\mathcal{M}_{e^-e^-\rightarrow e^-e^-}&= \frac{e^2}{(p_1^\alpha-p_3^\alpha)^2}[\bar{u}_3^{s_3}\gamma^\mu u_1^{s_1}][\bar{u}_4^{s_4}\gamma_\mu u_2^{s_2}]\nonumber\\
&-\frac{e^2}{(p_1^\alpha-p_4^\alpha)^2}[\bar{u}_3^{s_3}\gamma^\mu u_2^{s_2}][\bar{u}_4^{s_4}\gamma_\mu u_1^{s_1}],\label{amplitudes}
\end{align}
where
\begin{align*}	u_{1}^{\uparrow}=N\begin{pmatrix}
		1\\
		0\\
		\frac{p}{E+m}\\
		0
	\end{pmatrix},\quad u_{1}^{\downarrow}=N\begin{pmatrix}
		0\\
		1\\
		0\\
		\frac{-p}{E+m}
	\end{pmatrix},
\end{align*}
\vskip-0.3cm
\begin{align*}	u_{2}^{\uparrow}=N\begin{pmatrix}
 1 \\
 0 \\
 -\frac{p}{E+m} \\
 0 \\
\end{pmatrix},\quad u_{2}^{\downarrow}=N 
\begin{pmatrix}
 0 \\
 1 \\
 0 \\
 \frac{p}{E+m} \\
\end{pmatrix},
\end{align*}
\begin{equation*}
\bar{u}_{3}^{\uparrow}=N 
\begin{pmatrix}
 1, & 0, & \frac{p \cos\theta}{E+m}, & \frac{p e^{-i \phi}\sin\theta}{E+m} \\
\end{pmatrix}
\cdot\gamma^{0},
\end{equation*}
\begin{equation*}
\bar{u}_{3}^{\downarrow}=N 
\begin{pmatrix}
 0, & 1, & \frac{p e^{i\phi}\sin\theta}{E+m}, & \frac{-p\cos\theta}{E+m} \\
\end{pmatrix}
\cdot\gamma^{0},
\end{equation*}
\begin{equation*}
\bar{u}_{4}^{\uparrow}=N 
\begin{pmatrix}
 1, & 0, & \frac{-p \cos\theta}{E+m}, & \frac{-p e^{-i\phi}\sin\theta}{E+m} \\
\end{pmatrix}
\cdot\gamma^{0},
\end{equation*}
\begin{equation}
\bar{u}_{4}^{\downarrow}=N 
\begin{pmatrix}
 0, & 1, & \frac{-p e^{i \phi} \sin\theta}{E+m}, & \frac{p \cos\theta}{E+m} \\
\end{pmatrix}
\cdot\gamma^{0},\label{spinors}
\end{equation}
and the normalization $N=\sqrt{E+m}$ in terms of the mass of the electron $m$.

\subsection{Wigner rotations}
From special relativity, the coordinates between two different inertial frames are related by
\begin{equation}
x^{'\mu}\equiv T(\Lambda,b)x^{\mu}=\Lambda^{\mu}_{\ \nu}x^{\nu}+b^{\nu},
\end{equation}
where $T(\Lambda,b)$ represents a Lorentz transformation given by a boost $\Lambda$ and a translation $b$. This transformation induces a linear unitary transformation on vectors in the Hilbert space via $\ket{\psi}\rightarrow U(\Lambda,b)\ket{\psi}$, and for the irreducible unitary representation $\ket{p,s}$, this means a linear combination 
\begin{equation}
\ket{p,s}\rightarrow U(\Lambda,b)\ket{p,s}=\sum_{\sigma'}C_{\sigma'\sigma}(\Lambda,p)\ket{p^{\Lambda},\sigma'}.\label{linearcombination}
\end{equation}
For massive particles, it is possible to choose some standard 4-momentum $k^{\mu}$ (usually taken in the particle's rest frame) such that, under a standard Lorentz transformation $L(p)$, takes $k^{\mu}$ to $ p^{\mu}$. In the Hilbert space, it is 
\begin{equation}
\ket{p,s}\equiv U(L(p))\ket{k,s'}.
\end{equation}
Applying  $U(\Lambda,b=0)$ to both sides and using the composition property $U(\Lambda)U(L(p))=U(\Lambda L(p))$, one can insert the identity $L(\Lambda p)L^{-1}(\Lambda p)$ to obtain
\begin{align}
U(\Lambda)\ket{p,s}\equiv U(L(\Lambda p))U(W(\Lambda,p))\ket{k,s'}.
\end{align}
$W(\Lambda,p)\equiv L^{-1}(\Lambda p)\Lambda L(p)$ is the 
\textit{Wigner rotation} \cite{wigner}, since $L(p)$ takes the standard momentum $k$ to $L(p)k=p$, $\Lambda$ maps $p$ to $\Lambda p$, and $L^{-1}(\Lambda p)$ takes $\Lambda p$ back to $k$. Then, $W(\Lambda,p)$ is an element of the \textit{Wigner little group}, defined as the subgroup of the homogeneous Lorentz group that leaves  $k^{\mu}$ invariant, $W^{\mu}_{\ \nu}k^{\nu}=k^{\mu}$. This makes it possible to express Eq. \eqref{linearcombination} as
\begin{equation}
U(\Lambda,b)\ket{p,s}=\sum_{s'}D_{s's}^{\pm}(W(\Lambda,p))\ket{p^{\Lambda},s'},\label{wignerrotation}
\end{equation}
with $D_{s's}^{\pm}(W(\Lambda,p))$ being a suitable representation of the Wigner little group, which for the spin-1/2 case is just SU(2). Its matrix form is
\begin{equation}
D_{s's}^{\pm}(W(\Lambda,p))=\begin{pmatrix}
\cos\dfrac{\Omega}{2} & \pm\sin\dfrac{\Omega}{2}\\\\
\mp\sin\dfrac{\Omega}{2} & \cos\dfrac{\Omega}{2}
\end{pmatrix},\label{D's}
\end{equation}
where the parameter $\Omega$ denotes the Wigner rotation angle which expresses the rotation of the spin induced by two non-collinear boosts in the directions of $\pm z$ and $-x$, characterized by two rapidities $\eta$ and $\omega$, respectively. This relation reads  
\begin{equation}
\tan\Omega=\frac{\sinh\eta\sinh\omega}{\cosh\eta+\cosh\omega},\label{omegatheta}
\end{equation}
with $\eta,\omega\in\mathbb{R}$. The Wigner angle $\Omega$ vanishes whenever either rapidity is zero, and its sign is determined by the product $\sinh\eta\sinh\omega$, since $\cosh\eta+\cosh\omega>0$.

\section{zeroth- and first- Order Approximation of Feynman Amplitudes}\label{aproximation}
An specific interaction can be found by the expression
\begin{equation}
V=\frac{1}{(2\pi)^{3}}\int\mathcal{M}e^{i\boldsymbol{k}\cdot\boldsymbol{x}}d^{3}\boldsymbol{k},\label{potential}
\end{equation} 
where the scattering amplitude $\mathcal{M}$ depends on all final and initial momentum $\boldsymbol{k}=\boldsymbol{p}_{f}-\boldsymbol{p}_{i}$, constrained by energy-momentum conservation. Since each $\mathcal{M}$ is made by the product of therms  $\bar{u}_{f}\gamma^{\mu}u_{i}$, it is possible to decompose each $\gamma$
to obtain the zeroth-, first-, and higher-order terms in momentum, with the purpose of finding the non-relativistic and relativistic approximations of the Feynman amplitudes. These expressions for the specific product $\bar{u}_{3}^{s_{3}}\gamma^{\mu}u_{1}^{s_{1}}$ are given by \cite{sakurai}  
\begin{align*}
\bar{u}_{3}^{s_{3}}\gamma^{0}u_{1}^{s_{1}}=\xi^{s_{3}\dagger}\Bigg[1+\frac{\boldsymbol{p}_{3}\cdot\boldsymbol{p}_{1}}{(2m)^{2}}+\frac{i\boldsymbol{\sigma}\cdot(\boldsymbol{p}_{3}\times\boldsymbol{p}_{1})}{(2m)^{2}}\Bigg]\xi^{s_{1}}
\end{align*}
\begin{equation}
\text{or}\quad\quad\bar{u}_{3}^{s_{3}}\gamma^{0}u_{1}^{s_{1}}=\xi^{s_{3}\dagger}[1+\mathcal{O}(\boldsymbol{p}^{2})]\xi^{s_{1}},\label{gamazero}
\end{equation}
\begin{align*}
\bar{u}_{3}^{s_{3}}\gamma^{i}u_{1}^{s_{1}}=\xi^{s_{3}\dagger}\Bigg[\frac{-i(\boldsymbol{p}_{3}+\boldsymbol{p}_{1})}{2m}+\frac{\boldsymbol{\sigma}\times(\boldsymbol{p}_{3}-\boldsymbol{p}_{1})}{2m}\Bigg]\xi^{s_{1}}
\end{align*}
\begin{equation}
\text{or}\quad\quad\bar{u}_{3}^{s_{3}}\gamma^{i}u_{1}^{s_{1}}=\xi^{s_{3}\dagger}\mathcal{O}(\boldsymbol{p})\xi^{s_{1}},\label{gamai}
\end{equation}
where $\xi^{\uparrow}=(1,0)^{\text{T}}$ and $\xi^{\downarrow}=(0,1)^{\text{T}}$ represent spin polarizations along the particle’s momentum direction; i.e, the helicity. At very low energies, the first term of Eq. \eqref{gamazero} gives the \textbf{Coulomb interaction} $e^{2}/4\pi r$. In the next order of approximation, the momentum has acquired a more significant value and gives rise to other interactions through the first term (current term) and the second term (dipole term) of Eq. \eqref{gamai}. These kinds of interactions are specifically the \textbf{current-current}, \textbf{dipole-dipole}, and \textbf{current-dipole} interactions. For transitions that conserve the spin projection, the current terms are the same for $u_{1}^{\uparrow}\rightarrow u_{3}^{\uparrow}$ and $u_{1}^{\downarrow}\rightarrow u_{3}^{\downarrow}$ since there is no distinction between $(1,0)\cdot(1,0)^{\text{T}}$ and $(0,1)\cdot(0,1)^{\text{T}}$. However, this is not the case for the dipole terms, since $(1,0)\sigma_{z}(1,0)^{\text{T}}=-(0,1)\sigma_{z}(0,1)^{\text{T}}$, which appear in terms involving $\gamma^{1}$ and $\gamma^{2}$. The matrices $\sigma_{x}$ and $\sigma_{y}$ do not contribute to transitions that conserve the spin projection, which implies that the  \textit{z-}component term involving $\gamma^{3}$ vanishes. Explicitly, the previous considerations lead to
\begin{align}
\bar{u}_{3}^{\uparrow}\gamma^{1}u_{1}^{\uparrow}=\frac{-ip}{2m}\sin\theta\cos\phi-\frac{p}{2m}\sin\theta\sin\phi,\label{gamma1up}
\end{align}
\begin{align}
\bar{u}_{3}^{\uparrow}\gamma^{2}u_{1}^{\uparrow}=\frac{-ip}{2m}\sin\theta\sin\phi+\frac{p}{2m}\sin\theta\cos\phi,\label{gamma2up}
\end{align}
\begin{align}
\bar{u}_{3}^{\uparrow}\gamma^{3}u_{1}^{\uparrow}=\bar{u}_{3}^{\downarrow}\gamma^{3}u_{1}^{\downarrow}=\frac{-i(p\cos\theta+p)}{2m}+0;\label{spinsame}
\end{align}
the zero was left evident to symbolize that there is no dipole term in that product. The values of $\bar{u}_{3}^{\uparrow}\gamma^{1(2)}u_{1}^{\uparrow}$ are repeated for $\bar{u}_{3}^{\downarrow}\gamma^{1(2)}u_{1}^{\downarrow}$ by making the substitution $p\rightarrow-p$ in the second terms of Eqs \eqref{gamma1up} and \eqref{gamma2up}, respectively. For the remaining transitions of the \textit{t} and \textit{u} channels, where spin projection is maintained before and after scattering ($\uparrow\rightarrow\uparrow$ or $\downarrow\rightarrow\downarrow$), the change of sign in $p$ must be considered as follows: substitution $p\rightarrow-p$ in Eqs. \eqref{gamma1up}-\eqref{spinsame} ensures the values of $\bar{u}_{4}^{\uparrow(\downarrow)}\gamma^{i}u_{2}^{\uparrow(\downarrow)}$; this happens because the problem is being analyzed from the center-of-mass frame. For $\bar{u}_{4}^{\uparrow(\downarrow)}\gamma^{i}u_{1}^{\uparrow(\downarrow)}$ and $\bar{u}_{3}^{\uparrow(\downarrow)}\gamma^{i}u_{2}^{\uparrow(\downarrow)}$, presented in the \textit{u} channel, the substitution $p\rightarrow-p$ in the only non-zero value of the component $\gamma^{3}$ allows one to obtain a distinct result relative to the \textit{t-}channel transitions; this is due to the fact that contributions from the initial particle momenta arise exclusively in the \textit{z-}direction. Furthermore, along with the exchange of these initial electrons comes the exchange of their dynamic variable $p$. On the other hand, for transitions in which the spin projection flips, the contributions arise solely from the dipole terms, since the current terms do not mediate spin-flip transitions. In this case, only matrices $\sigma_{x}$ and $\sigma_{y}$ contribute. Moreover, the relations $(1,0)\sigma_{x}(0,1)^{\text{T}}=(0,1)\sigma_{x}(1,0)^{\text{T}}$ and $(1,0)\sigma_{y}(0,1)^{\text{T}}=-(0,1)\sigma_{y}(1,0)^{\text{T}}$,
which appear in terms involving the gamma matrices, allow one to determine the corresponding spin-flip products
\begin{align}
\bar{u}_{3}^{\downarrow}\gamma^{1}u_{1}^{\uparrow}=0+\frac{i(p\cos\theta-p)}{2m},\label{jj}
\end{align}
\begin{align}
\bar{u}_{3}^{\downarrow}\gamma^{2}u_{1}^{\uparrow}=\bar{u}_{3}^{\uparrow}\gamma^{2}u_{1}^{\downarrow}=0-\frac{(p\cos\theta-p)}{2m},\label{gamma2du}
\end{align}
\begin{align}
\bar{u}_{3}^{\downarrow}\gamma^{3}u_{1}^{\uparrow}=0-\frac{ip}{2m}\sin\theta e^{i\phi};\label{spinflips}
\end{align}
again, the zero was left in evidence to symbolize that there is no contribution from current terms. The terms in $\bar{u}_{3}^{\downarrow}\gamma^{1(3)}u_{1}^{\uparrow}$ are repeated for $\bar{u}_{3}^{\uparrow}\gamma^{1(3)}u_{1}^{\downarrow}$ by making the substitution $i\rightarrow-i$ in the second terms of Eqs. \eqref{jj} and \eqref{spinflips}. The considerations discussed above for the other transitions, which also apply to the \textit{t-} and \textit{u-}channels, are taken into account to compute the total scattering amplitude $\mathcal{M}=\mathcal{M}_{t}-\mathcal{M}_{u}$. The amplitudes corresponding to the interactions of Coulomb $\mathcal{M}^{Co}$, current-current $\mathcal{M}^{CC}$, dipole-dipole $\mathcal{M}^{DD}$, and current-dipole $\mathcal{M}^{CD}$ are explicitly written, respectively, in Eqs. \eqref{Coulomb}-\eqref{CD} in the Appendix.

\section{$\boldsymbol{e^{-}e^{-}}\rightarrow\boldsymbol{e^{-}e^{-}}$ }\label{2-2}\label{2-2}
Consider the collision shown in Fig. \ref{fig:Moller general}. First, we analyze the initial entanglement between both particles, represented by the blue line. We then consider a separable state analyzed both in the center-of-mass frame and under a perpendicular boost.
\subsection{Entangled initial state}\label{entangled}
Suppose the initial spin-entangled state for two interacting electrons in a center-of-mass frame
\begin{align}
\ket{\psi_{\text{in}}}=\frac{1}{\sqrt{2}}[\ket{\textup{\textbf{p}}_{1},\uparrow}\otimes\ket{\textup{\textbf{p}}_{2},\downarrow}+\ket{\textup{\textbf{p}}_{1},\downarrow}\otimes\ket{\textup{\textbf{p}}_{2},\uparrow}],
\label{estadoinicialE}
\end{align}
where the product of $\ket{\Psi^{+}}$ with the momentum states was included using Eq. \eqref{in-out kets}. 
The transition state, constructed from the scattering amplitudes applying the operator $\widehat{\mathcal{T}}$, is \label{gamma1du} 
\begin{align}
\ket{\psi_{\text{trans}}}&=i\SumInt_{ \textup{\textbf{p}}_{3},\textup{\textbf{p}}_{4},r,s}(2\pi)\delta(E_{\textbf{\text{p}}_{1}}+E_{\textbf{\text{p}}_{2}}-E_{\textbf{\text{p}}_{3}}-E_{\textbf{\text{p}}_{4}})\nonumber\\
&(2\pi)^{3}\delta^{3}(\textbf{\text{p}}_{1}+\textbf{\text{p}}_{2}-\textbf{\text{p}}_{3}-\textbf{\text{p}}_{4})\frac{1}{\sqrt{2}}\Big[\mathcal{M}(\uparrow\downarrow\rightarrow rs)\nonumber\\
&+\mathcal{M}(\downarrow\uparrow\rightarrow rs)\Big]\ket{\textbf{\text{p}}_{3},r}\otimes\ket{\textbf{\text{p}}_{4},s};\label{estadoex1E}
\end{align}
here, the indices expressing transitions $\textbf{\text{p}}_{1}(\textbf{\text{p}}_{2})\rightarrow\textbf{\text{p}}_{3}(\textbf{\text{p}}_{4})$ are omitted in the $\mathcal{M}'s$ but transitions for the final spin possibilities, $rs$, remain. The density matrix is built from $\ket{\psi_{\text{out}}}=\ket{\psi_{\text{in}}}+\ket{\psi_{\text{trans}}}$, and the particular spin state of one electron is obtained through Eq. \eqref{matricesreducidas}, as
\begin{align}
\rho_{e^{-}_{1}}=\text{Tr}&_{\textbf{p}_{e^{-}_{1}}}\text{Tr}_{e^{-}_{2}}\nonumber\\
[\mathcal{N}^{-1}(\ket{\psi_{\text{in}}}+\ket{\psi_{\text{trans}}})&(\bra{\psi_{\text{in}}})+\bra{\psi_{\text{trans}}})]\label{matrizreducida}. 
\end{align}
The normalization factor obtained by Eq. \eqref{TotalN}, divided by $(2EV)^{2}$, is given by
\begin{align}
\mathcal{N}&=1+\frac{T}{2V}\frac{|\textbf{p}_{3}|}{E}\frac{1}{(8\pi E)^{2}}\sum_{r,s}\int d\Omega\nonumber\\
|\mathcal{M}(\uparrow\downarrow&\rightarrow rs)|^{2}+\mathcal{M}(\uparrow\downarrow\rightarrow rs)\bar{\mathcal{M}}(\downarrow\uparrow\rightarrow rs),\label{normalizationE}
\end{align}
where $T$ is the time duration such that $2 \pi \delta (E_\text{f} - E_\text{i}) =\lim_{T\rightarrow \infty} \int_{-T/2}^{+T/2} \, dt \, \exp[i (E_\text{f} - E_\text{i}) t]$, and $V = (2 \pi)^3 \delta^3 ({\boldsymbol\epsilon})$ with ${\boldsymbol\epsilon} \rightarrow \bf{0}$ is the volume of space. The energy of the colliding particles $E$ is measured in the center-of-mass frame and $|\textbf{p}_{3}|=\sqrt{E^{2}-m^{2}}$. A barred quantity means the Hermitian conjugation of $\mathcal{M}$ and $d\Omega = \sin \theta d \theta d \phi$ is the infinitesimal solid angle. The first term corresponds to the normalization of the free part $\mathcal{N}_{\text{in}}$, while the second corresponds to $\mathcal{N}_{\text{trans}}$. For the crossed terms $\mathcal{N}_{\text{cross}}$, it is recalled that at the tree level these terms vanish. In the calculation of $\rho_{e^{-}_{1}}$, the terms that arise from $\ket{\psi_{\text{in}}}\bra{\psi_{\text{in}}}$ are easy to calculate, since tracing out the other electron makes $\rho_{\text{in}}$ diagonal with eigenvalues equal to $1/2$. This result is expected whenever a subsystem initially entangled with another is reduced by tracing out its partner, thereby yielding a mixed state. For the transition contribution, $\ket{\psi_{\text{trans}}}\bra{\psi_{\text{trans}}}$, we obtain the following
\begin{align}
\rho&_{\text{trans}}=\frac{T}{4V}\frac{|\textbf{p}_{3}|}{E}\frac{1}{(8\pi E)^{2}}\sum_{rsl}\int d\Omega[\mathcal{M}(\uparrow\downarrow\rightarrow rs)\nonumber\\
+&\mathcal{M}(\downarrow\uparrow\rightarrow rs)][\bar{\mathcal{M}}(ls\rightarrow \uparrow\downarrow)+\bar{\mathcal{M}}(ls\rightarrow\downarrow\uparrow)]\ket{r}\bra{l}.   
\end{align}
After evaluating all the amplitudes $\mathcal{M}'s$ through the Eqs. \eqref{Coulomb}-\eqref{CD}, it is find that the density matrix does not change after the scattering; i.e.,
\begin{align}
\rho_{\text{in}}=\begin{pmatrix}
\frac{1}{2} & 0 \\
0 & \frac{1}{2}
\end{pmatrix}\quad\quad\xrightarrow{\text{Scattering}}\quad\quad\rho_{\text{out}}=\begin{pmatrix}
\frac{1}{2} & 0 \\
0 & \frac{1}{2}
\end{pmatrix}.\label{ghzall}
\end{align}
This result suggests that, under the interactions considered in this work, scattering cannot generate additional correlations when the system is initially maximally entangled. This can be understood from the fact that no further mixedness can be introduced into a reduced system whose mixedness is already maximal prior to the process.

\subsection{Unentagled initial state}\label{unentagled}
\begin{figure}[H]
\captionsetup{justification=centering}
\centering
\includegraphics[scale=0.23]{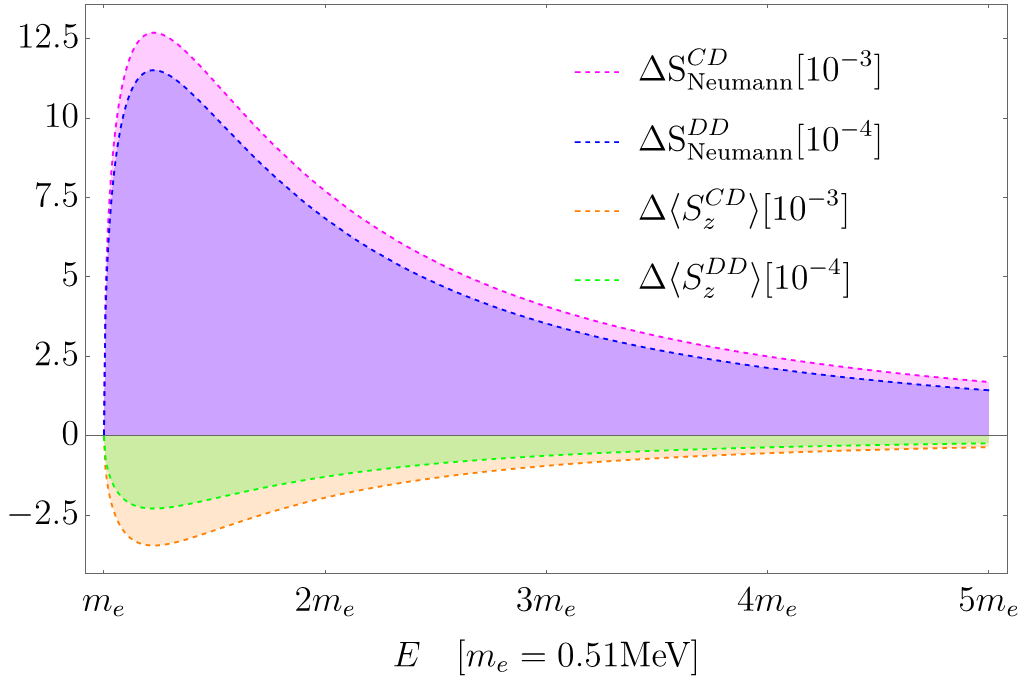} 
\caption{Variation of the von Neumann entropies and spin expectation values in the $z$ direction of one electron as functions of the collision energy for the current–dipole and dipole–dipole interactions.}
\label{fig:todas}
\end{figure}

Consider now an initial separable state given by
\begin{align}
\begin{matrix}
\ket{\psi_{\text{in}}}=\ket{\textup{\textbf{p}}_{1},\uparrow}\otimes\ket{\textup{\textbf{p}}_{2},\uparrow}
\end{matrix},\label{estadoinicialNE}
\end{align}
with the transition state
\begin{align}
&\ket{\psi_{\text{trans}}}=i\SumInt_{ \textup{\textbf{p}}_{3},\textup{\textbf{p}}_{4},r,s}(2\pi)\delta(E_{\textbf{p}_{1}}+E_{\textbf{p}_{2}}-E_{\textbf{p}_{3}}-E_{\textbf{p}_{4}})(2\pi)^{3}\nonumber\\
&\delta^{3}(\textbf{p}_{1}+\textbf{p}_{2}-\textbf{p}_{3}-\textbf{p}_{4})\mathcal{M}(\uparrow\uparrow\rightarrow rs)\ket{\textbf{p}_{3},r}\otimes\ket{\textbf{p}_{4},s}.\label{estadoex1NE}
\end{align}
The same considerations on the notation made in the previous part are repeated here. The normalization for the reduced density matrix $\rho_{e^{-}_{1}}$, again divided by $(2EV)^{2}$, is 
\begin{align}
\mathcal{N}=1+\frac{T}{2V}\frac{|\textbf{p}_{3}|}{E}\frac{1}{(8\pi E)^{2}}\sum_{r,s}\int d\Omega|\mathcal{M}(\uparrow\uparrow\rightarrow rs)|^{2}.\label{normalizationNE}
\end{align}
Note that only one scattering amplitude $\mathcal{M}$ appears given the initial configuration. The initial matrix $\rho_{\text{in}}$ is composed of only one element $\rho_{\text{in}}^{11}=1$ and $0$ for any other $\rho^{ij}$; that is, the matrix that characterizes a pure state. The transition matrix $\rho_{\text{trans}}$ is proportional to the product of the single amplitude and its complex conjugate, as
\begin{align}
\rho_{\text{trans}}=\frac{T}{2V}&\frac{|\textbf{p}_{3}|}{E}\frac{1}{(8\pi E)^{2}}\sum_{rsl}\int d\Omega [\mathcal{M}(\uparrow\uparrow\rightarrow rs)]\nonumber\\
&\times[\bar{\mathcal{M}}(ls\rightarrow\uparrow\uparrow)]\ket{r}\bra{l}.\label{rhotransA}
\end{align}
Upon integrating the squared amplitudes over the full solid angle, divergences arise due to the singular behavior of the integrand in the forward and backward limits  $\theta\rightarrow0$ and $\theta\rightarrow\pi$. The divergence arises from the presence of the factors $(1-\cos\theta)$ and $(1+\cos\theta)$ in the denominators of the \textit{t-} and \textit{u-}channels, respectively. This problem is also present when integrating the differential cross section $d\sigma/d\Omega$. Therefore, in order to investigate the modifications in the elements of the density matrix after scattering, we perform the angular integration over the finite interval $0.05\leq\theta\leq3.1$, excluding the singular endpoints $\theta=0$ and $\theta=\pi$. In principle, if these modifications occur, mediated by Eq. \eqref{rhotransA}, the von Neumann entropy
\begin{equation}
\text{S}_{\text{Neumann}} =-\text{Tr} [\rho_{e^{-}_{1}} \ln \rho_{e^{-}_{1}}],\label{sneumann}
\end{equation}
It would exhibit a variation in $\Delta\text{S}_{\text{Neumann}}>0$ before and after scattering. In that case, we may infer dynamical information by an experimental observable such as the variation of spin expectation value along a given direction, defined by 
\begin{eqnarray}
&&\Delta\langle S_{x,y,z}\rangle = \langle S_{x,y,z}\rangle_{\mathrm{out}} - \langle S_{x,y,z}\rangle_{\mathrm{in}}\nonumber \\
&& =\frac{1}{2} \text{Tr}[\sigma_{x,y,z} \,\, \rho_{e_{1}^{-}}] -\frac{1}{2} \text{Tr}[\sigma_{x,y,z} \,\, \rho_{\text{in}}].\label{variation}
\end{eqnarray}
Evaluating all the amplitudes described in Eqs.  \eqref{Coulomb}-\eqref{CD}, Fig. \ref{fig:todas} shows the contributions made by all interactions. 
As can be observed, the current-dipole and dipole-dipole interactions contribute to the variation of Eqs. \eqref{sneumann} and \eqref{variation}. This is equivalent to the generation of nonvanishing diagonal entries in the density matrix, which indicates the emergence of correlations between the two particles. In the absence of a dipole term in Eq. \eqref{spinsame}, the current–dipole contribution becomes dominant by a factor of 10 compared to the dipole–dipole interaction. This is directly related to the fact that the current expression does not allow spin flipping, while the dipole expression does. The presence of diagonal terms is a consequence of the symmetry of the amplitudes that contribute to $\rho_{\text{trans}}^{11}$ and  $\rho_{\text{trans}}^{22}$, 
together with the intrinsic asymmetry of the initial density matrix, for which $\rho^{11}_{\text{in}}=1$ and $\rho^{22}_{\text{in}}=0$. The absence of values in the non-diagonal terms implies that the state cannot become purer than it started. This absence  becomes manifest due to the constraints imposed by the interaction encoded in the amplitudes Eqs. \eqref{Coulomb}-\eqref{CD}; i.e., the amplitudes of those off-diagonal elements are zero. 

\begin{figure}[H]
\captionsetup{justification=centering}
\centering
\includegraphics[scale=0.19]{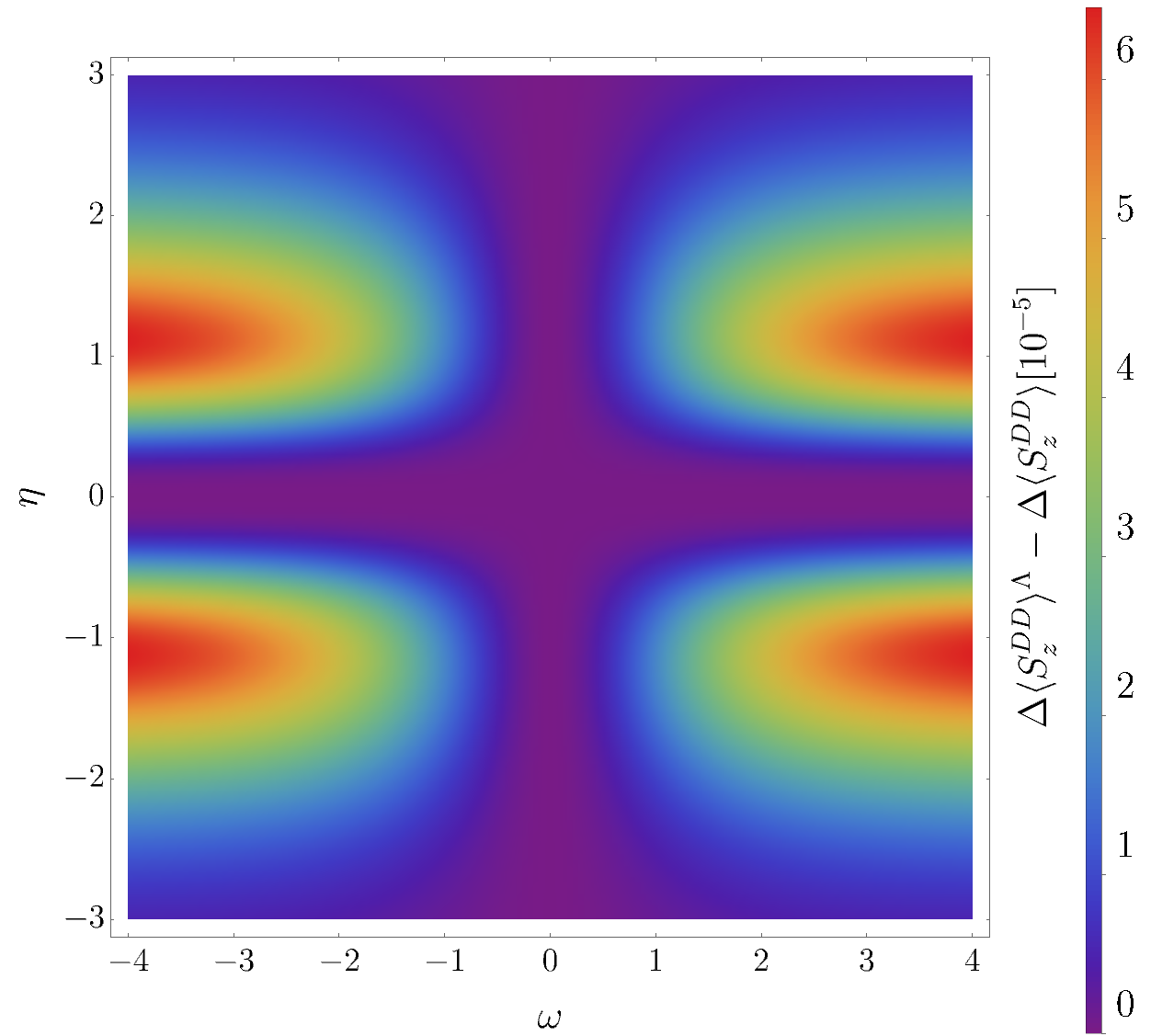} 
\caption{Variation of the spin expectacion values in the $z$ direction of one electron as functions of the rapidities for the dipole-dipole interaction.}
\label{fig:VzDD}
\end{figure}

\begin{figure}[H]
\captionsetup{justification=centering}
\centering
\includegraphics[scale=0.19]{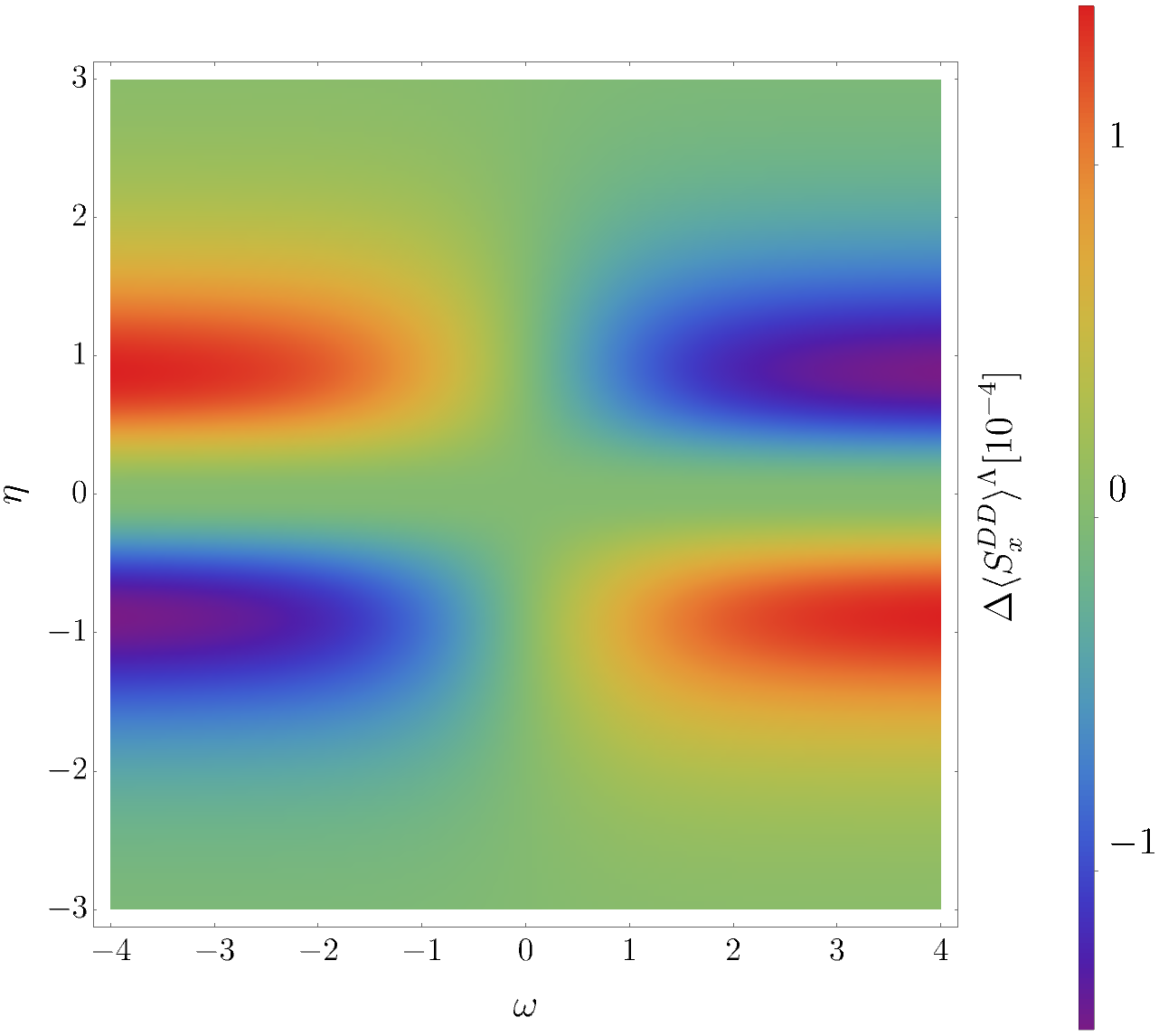} 
\caption{Variation of the spin expectacion value in the $x$ direction of one electron as functions of the rapidities for the dipole-dipole interaction.}
\label{fig:VxDD}
\end{figure}

Now, let us consider the case in which an observer moves in the $\pm x$ direction. Using Eq. \eqref{wignerrotation}, the initial state in that new frame of reference becomes
\begin{align}
\ket{\psi^{\Lambda}_{\text{in}}}=\ket{\textup{\textbf{p}}_{1}^{\Lambda}}&\otimes\ket{\textup{\textbf{p}}_{2}^{\Lambda}}\Big(\cos^{2}\frac{\Omega}{2}\ket{\uparrow\uparrow}-\sin^{2}\frac{\Omega}{2}\ket{\downarrow\downarrow}\nonumber\\
&-\sin\frac{\Omega}{2}\cos\frac{\Omega}{2}[\ket{\uparrow\downarrow}-\ket{\downarrow\uparrow}]\Big),
\end{align}
where $\Omega$ is given by Eq. \eqref{omegatheta}. For the transition state, the Lorenz invariance  
\begin{equation}
d\Pi=\int_{\textup{\textbf{p}}_{3},\textup{\textbf{p}}_{4}}(2\pi)^{4}\delta^{4}(p_{\text{i}}^{\alpha(1)}+p_{\text{i}}^{\alpha(2)}-p_{\text{f}}^{\alpha(1)}-p_{\text{f}}^{\alpha(2)})
\end{equation} allows us to obtain
\begin{align}
\ket{\psi_{\text{trans}}^{\Lambda}}=i&\SumInt_{ \textbf{p}_{3},\textbf{p}_{4},r,s}(2\pi)\delta(E_{\textbf{p}_{1}}+E_{\textbf{p}_{2}}-E_{\textbf{p}_{3}}-E_{\textbf{p}_{4}})\nonumber\\
(2&\pi)^{3}\delta^{3}(\textbf{p}_{1}+\textbf{p}_{2}-\textbf{p}_{3}-\textbf{p}_{4})\mathcal{M}(\uparrow\uparrow\rightarrow rs)\nonumber\\
&\sum_{m}D^{-}_{mr}\ket{\textbf{p}_{3}^{\Lambda},m}\otimes\sum_{n}D^{+}_{ns}\ket{\textbf{p}_{4}^{\Lambda},n},    
\end{align}
where $D^{\pm}_{ij}$ are given by Eq. \eqref{D's}. Note that if $\omega\rightarrow0$, $\Omega\rightarrow0$, and the states $\ket{\psi_{\text{in}}}$ and $\ket{\psi_{\text{trans}}}$ described in the center-of-mass frame are recovered. The other invariances, $2E_{\textbf{p}^{\Lambda}}\delta^{3}(\textbf{p}^{\Lambda}-\textbf{p}'^{\Lambda})=2E_{\textbf{p}}\delta^{3}(\textbf{p}-\textbf{p}')$ and $T_{\textbf{p}^{\Lambda}}V_{\textbf{p}^{\Lambda}}=T_{\textbf{p}}V_{\textbf{p}}$, together with the same mechanism used in the center-of-mass reference frame, lead to $\mathcal{N}^{\Lambda}=\mathcal{N}$; i.e., 
\begin{align}
\mathcal{N}^{\Lambda}=1+\frac{T}{2V}\frac{|\textbf{p}_{3}|}{E}\frac{1}{(8\pi E)^{2}}\sum_{r,s}\int d\Omega|\mathcal{M}(\uparrow\uparrow\rightarrow rs)|^{2}.
\end{align}
This is an expected result, since the trace of the density matrix, which is related to probability conservation, must remain invariant under changes of reference frame. Mathematically, this can be expressed as $\text{Tr}(D\rho D^{\dagger})=\text{Tr}(D^{\dagger}D\rho)=\text{Tr}(\rho)$, where the cyclic property of the trace and the unitarity of $D^{\pm}$ have been used. The initial matrix continues to characterize a pure state, although elements depending on the angle $\Omega$ emerge throughout all rows and columns. However, the only term in which the cosine function appears independently is $\rho_{\text{in}}^{11\Lambda}=\cos^{4}(\Omega/2)+1/4\sin^{4}(\Omega)$, while the other elements are a function of $\sin(\Omega)$. This guaranties that, setting $\Omega=0$, the initial matrix described in the center-of-mass frame is recovered ($\rho_{\text{in}}^{11}=1$ and $0$ for any other $\rho^{ij}$). Furthermore, this is supported by the fact that 
\begin{equation}
\text{S}_{\text{Neumann}}(\rho_{\text{in}})=\text{S}_{\text{Neumann}}(\rho_{\text{in}}^{\Lambda})=0.
\end{equation}
To obtain the transition matrix $\rho_{\text{trans}}$, the same conditions and invariances appearing in $\mathcal{N}^{\Lambda}$ are taken into account, and, as in the center-of-mass frame, the off-diagonal elements vanish, resulting in the simplest form of the density matrix
\begin{align}
&\rho^{11\Lambda}_{\text{trans}}=\rho^{22\Lambda}_{\text{trans}}=|\mathcal{M}(\uparrow\uparrow\rightarrow\uparrow\downarrow)|^{2}+\nonumber\\
\sin^{4}\frac{\Omega}{2}&|\mathcal{M}(\uparrow\uparrow\rightarrow\uparrow\uparrow)|^{2}+\cos^{4}\frac{\Omega}{2}|\mathcal{M}(\uparrow\uparrow\rightarrow\downarrow\downarrow)|^{2}+\nonumber\\
\frac{1}{4}\sin^{2}&\Omega[|\mathcal{M}(\uparrow\uparrow\rightarrow\uparrow\uparrow)|^{2}+|\mathcal{M}(\uparrow\uparrow\rightarrow\downarrow\downarrow)|^{2}].
\end{align}
However, interestingly, the disappearance of the off-diagonal elements here is not due to the constraints imposed by the considered interactions, since their corresponding amplitudes are nonzero (as it does in the center-of-mass frame), but rather to a superposition between opposite contributions mediated by the factors accompanying the Wigner rotations. Thus, for example, one would have 
\begin{align}
\rho^{12\Lambda}_{\text{trans}}&\propto\cos^{2}\frac{\Omega}{2}\mathcal{M}(\uparrow\uparrow\rightarrow\uparrow\uparrow)\bar{\mathcal{M}}(\uparrow\uparrow\rightarrow\uparrow\downarrow)+\nonumber\\
&\cos^{2}\frac{\Omega}{2}\mathcal{M}(\uparrow\uparrow\rightarrow\uparrow\uparrow)\bar{\mathcal{M}}(\uparrow\uparrow\rightarrow\uparrow\downarrow)+\nonumber\\
\frac{1}{2}\sin\Omega&(|\mathcal{M}(\uparrow\uparrow\rightarrow\uparrow\uparrow)|^{2}-|\mathcal{M}(\uparrow\uparrow\rightarrow\downarrow\downarrow)|^{2})=0,   
\end{align}
although all $\mathcal{M}'s\neq0$. Thus, it is observed that the creation of coherence is due to the contribution of the values present in $\rho_{\text{in}}^{\Lambda}$ rather than those of $\rho_{\text{trans}}^{\Lambda}$, as the latter vanish. This phenomenon is explained by the rotation of the spin caused by the perpendicular motion, while the scattering process itself remains the same in both inertial frames. This same feature is reflected in  equality $\text{S}_{\text{Neumann}}(\rho_{\text{out}})=\text{S}_{\text{Neumann}}(\rho_{\text{out}}^{\Lambda})$,
confirming that a local unitary transformation does not affect any measure of entanglement, establishing the total entropy as a Lorentz invariant. Despite this, the same is not true for the variations of the spin expectation values, since, as mentioned before, nonzero values arise due to Wigner rotations. For the interaction dipole-dipole, Fig. \ref{fig:VzDD} displays the difference between the variations of the spin expectation values in both inertial frames along the $z$ direction, while Fig. \ref{fig:VxDD} shows the emergence of spin expectation values along the $x$ direction. Here, the energies in the two inertial frames are parameterized by the rapidities $\eta$ and $\omega$ in the following way:
\begin{align}
E=m\cosh\eta,\quad E^{\Lambda}=E\cosh\omega.
\end{align}
As can be observed, the orders of magnitude along the $x-$axis are ten times larger than those along the $z-$axis. This implies a predominance in the generation of off-diagonal elements, while the increase in the diagonal elements where contributions due to the scattering process were already present is comparatively smaller. We see that these diagonal and off-diagonal contributions ultimately balance each other in such a way that the entanglement, quantified by $\text{S}_{\text{Neumann}}$, remains invariant in both inertial frames. On the other hand, the variation of the spin expectation value that emerges along the $x-$axis is sensitive to changes in both the  $\omega,\eta<0$ and $\omega,\eta>0$ regions, whereas this is not the case along the $z-$axis. 
\begin{figure}[H]
\captionsetup{justification=centering}
\centering
\includegraphics[scale=0.23]{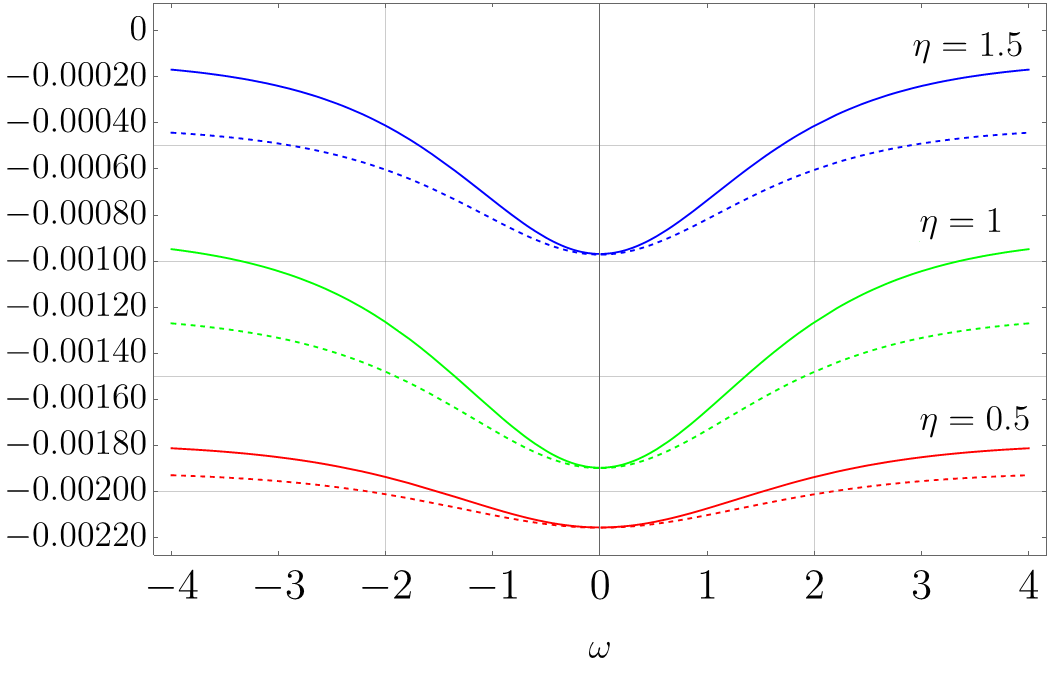} 
\caption{Variation of the spin expectation value in the $z$ direction for the current–dipole (solid line) and dipole–dipole (dashed line) interactions as a function of the rapidity $\omega$ for selected values of $\eta$.}
\label{fig:z}
\end{figure}
\begin{figure}[H]
\captionsetup{justification=centering}
\centering
\includegraphics[scale=0.23]{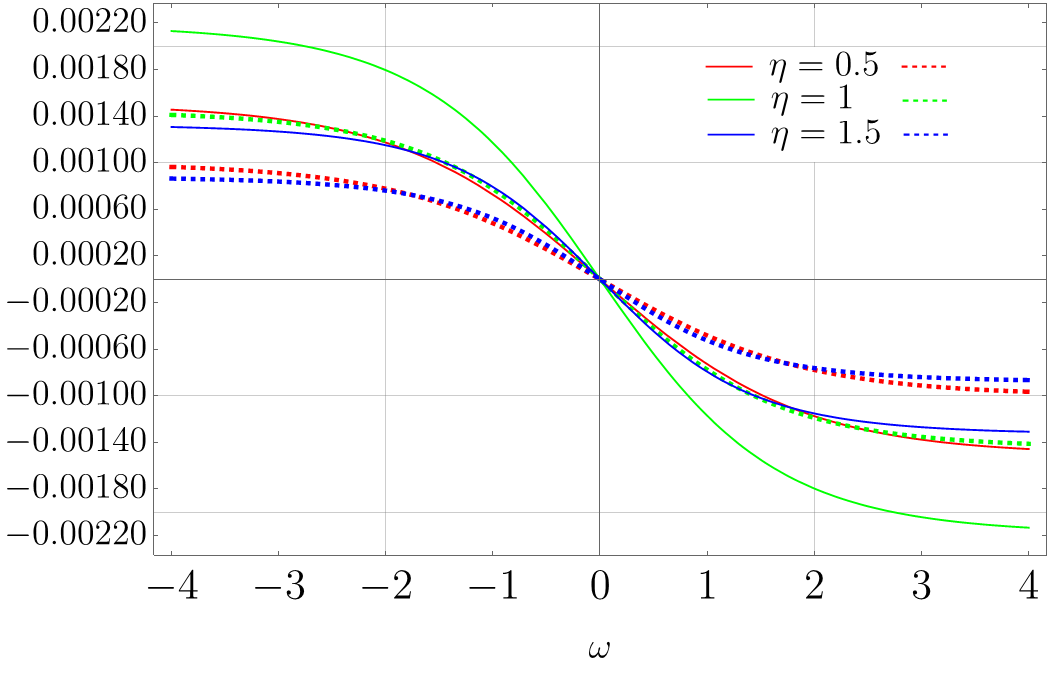} 
\caption{Variation of the spin expectation value in the $x$ direction for the current–dipole (solid line) and dipole–dipole (dashed line) interactions as a function of the rapidity $\omega$ for selected values of $\eta$.}
\label{fig:x}
\end{figure}
The results for the current–dipole interaction display behavior very similar to those of the dipole–dipole, with the exception that their orders of magnitude are approximately one order larger (see, for instance, Fig. \ref{fig:todas}). Therefore, it is more illustrative to compare both interactions for selected fixed collision energies measured in the center-of-mass frame; namely, $\eta=0.5$, $\eta=1$, and $\eta=1.5$. This is shown in Figs. \ref{fig:z} and \ref{fig:x}. In the first graph, the solid line corresponds to the current–dipole interaction, shifted along the $y-$axis to overlap with the dipole–dipole, which is represented by the dashed line and rescaled by a factor of 10. In the second plot, the solid and dashed lines continue to denote current-dipole and dipole–dipole interactions, respectively. In this case, however, the solid lines were not shifted, since they all share the same origin at zero. The dashed lines were again rescaled by a factor of 10. It can be seen that the difference between the corresponding interactions for $\eta=0.5$ and $\eta=1.5$ is very small, still indicating a growth towards a maximum near $\eta=1.2$, followed by a subsequent decrease. Interestingly, the dipole–dipole interaction for $\eta=1$ displays behavior remarkably similar to a tenth of the current–dipole interaction for $\eta=0.5$ and $\eta=1.5$. The same symmetry and antisymmetry displayed in Figs. \ref{fig:VzDD} and \ref{fig:VxDD}, respectively, are also observed here. Furthermore, as before, the values of these correlations increase with increasing rapidity $\omega$.

\section{$\boldsymbol{e^{-}e^{-}C}\rightarrow\boldsymbol{e^{-}e^{-}}C$ }\label{testemunha}
\begin{figure}[H]
\captionsetup{justification=centering}
\centering
\includegraphics[scale=0.24]{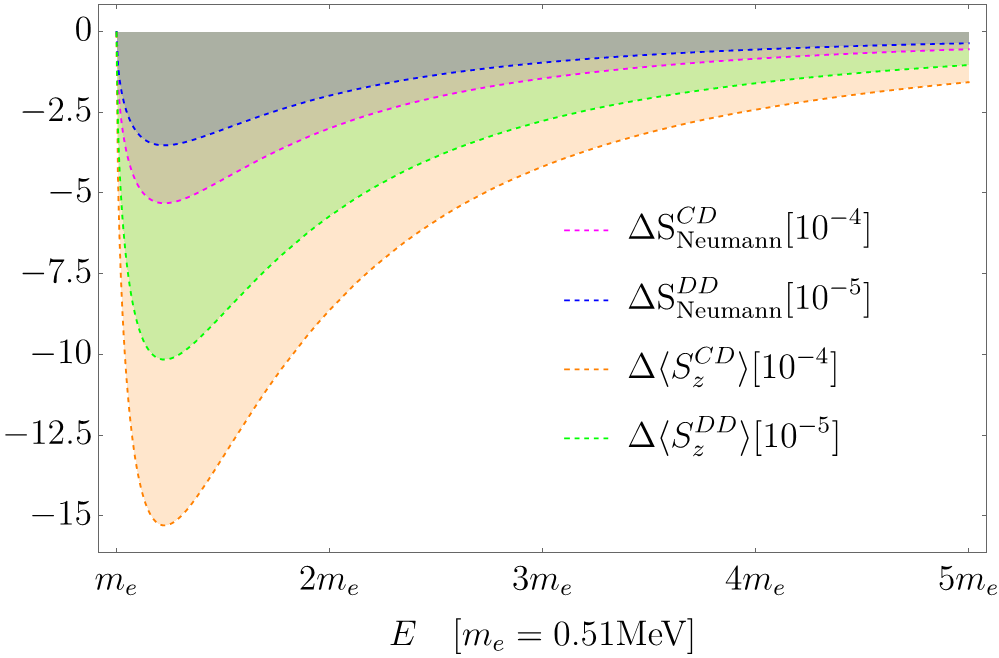} 
\caption{Variation of the von Neumann entropies and spin expectation values in the $z$ direction of particle $C$ as functions of the collision energy for the current–dipole and dipole–dipole interactions.}
\label{fig:Total}
\end{figure}

The entanglement between two interacting electrons with a witness particle $C$ through the $\ket{W}$ state is now considered as shown in Figs.  \ref{fig:Moller general} and \ref{fig:feyman} (red lines). From the results obtained in \cite{Fonseca}, the density matrix for $C$ 
when it is initially entangled with one or both interacting particles is described by 
\begin{align}
	\rho_{\text{C}}= \frac{1}{\mathcal{N}}\Big(\rho_{\text{C(in)}} + \rho_{\text{C(trans)}}\Big),
	\label{rhoC}
\end{align}
with
\begin{equation}
	\rho_{\text{C(in)}}=8E_{\textbf{\text{q}}}E^{2}V^{3} \times\label{cin}
\end{equation}
\begin{equation*}\times
	\begin{pmatrix}
	c_{1}^{2}+c_{3}^{2}+c_{5}^{2}+c_{7}^{2} & c_{1}c_{2}+c_{3}c_{4}+c_{5}c_{6}+c_{7}c_{8} \\\\
	c_{1}c_{2}+c_{3}c_{4}+c_{5}c_{6}+c_{7}c_{8} & c_{2}^{2}+c_{4}^{2}+c_{6}^{2}+c_{8}^{2} 
	\end{pmatrix}\label{rhoCin}
\end{equation*}
and 
\begin{equation}
	\rho_{\text{C(trans)}}=\frac{E_{\textbf{\text{q}}}TV^{2}|\textbf{\text{p}}_{3}|}{16\pi^{2}E}\SumInt_{s} d\Omega \times \label{rhoCtrans}
\end{equation}
\vskip-0.5cm
\begin{equation*}\times	
	\begin{pmatrix}
		\Lambda^{1,1}_{\text{AC}}(s)+\Lambda^{3,3}_{\text{AC}}(s) & \Lambda^{1,2}_{\text{AC}}(s)+\Lambda^{3,4}_{\text{AC}}(s) \\\\ 
		\Lambda^{1,2}_{\text{AC}}(s)+\Lambda^{3,4}_{\text{AC}}(s) & \Lambda^{2,2}_{\text{AC}}(s)+\Lambda^{4,4}_{\text{AC}}(s) 
	\end{pmatrix}.
\end{equation*}
The total initial state is constrained by $\sum_{j=1}^{8}c_{j}^{2}=1$, and $E_{\textbf{\text{q}}}$ represents the energy of $C$. The matrix elements $\Lambda^{i,j}_{\text{AC}}(s)$, present in Eq. \eqref{rhoCtrans}, are given by the expressions \ref{rhosA}. The total normalization was defined as
\bq
\mathcal{N} &=& \mathcal{N}_{\text{in}}+\mathcal{N}_{\text{trans}}
\nonumber \\
&=&8E_{\textbf{\text{q}}}E^{2}V^{3}+\frac{E_{\textbf{\text{q}}}TV^{2}|\textbf{\text{p}}_{3}|}{16\pi^{2}E}\SumInt_{r,s} d\Omega\Lambda(r,s),\label{explinorma}
\eq
and $\Lambda(r,s)$ stands for
\bq
\Lambda(r,s)&=&(c_{1}^{2}+c_{2}^{2}+c_{7}^{2}+c_{8}^{2})|\mathcal{M}(\uparrow\uparrow\rightarrow rs)|^{2} \nonumber\\
&+& (c_{3}^{2}+c_{4}^{2}+c_{5}^{2}+c_{6}^{2})|\mathcal{M}(\uparrow\downarrow\rightarrow rs)|^{2}\\
&+&2(c_{3}c_{5}+c_{4}c_{6})\mathcal{M}(\uparrow\downarrow\rightarrow rs) \mathcal{\bar{M}}(\downarrow\uparrow\rightarrow rs). \nonumber\label{lambdas}
\eq
Setting $c_4=c_6=c_7=(3)^{-1/2}$ and $c_i=0$ otherwise, the initial entangled state between the three particles $\ket{\text{W}}=(3)^{-1/2}[\ket{\downarrow\downarrow\uparrow}+\ket{\downarrow\uparrow\downarrow}+\ket{\uparrow\downarrow\downarrow}]$ is recovered. Research \cite{Fonseca} also showed that, at tree-level, the witness particle $C$ develops correlations with both electrons that depend on the collision energy $E$. An analogous result holds for other initial configurations compatible with the selected entangled state. However, it did not discuss which interactions, or even their nature, were responsible for the creation of those correlations. Here, the scattering amplitudes associated with each interaction are evaluated, and the variations of the von Neumann entropy and spin expectation value in $C$ are shown in Fig. \ref{fig:Total}. 
As in the previous section, the causes of the variations lie in current-dipole and dipole-dipole interactions. The values for this case are slightly smaller than those shown in Fig. \ref{fig:todas}, a feature that arises because the correlations must be distributed among three particles instead of two. Again, the absence of the dipole term in $\gamma^{3}$ of Eq. \eqref{spinsame} causes the current-dipole interaction to be greater than the dipole-dipole interaction. The symmetry and, consequently, the repetition of scattering amplitudes in the diagonal elements of the matrix, together with the vanishing of some others in the off-diagonal elements, cause this variation in the spin expectation value along the z-axis. Note that, in this case, the entropy variation for C decreases rather than increases, as occurs for the electron in the previous section. This is, in principle, allowed due to its initially unconstrained state; that is, a density matrix whose mixedness is neither maximal nor minimal, and which can therefore become either more pure or more mixed, respectively.
\section{Comments for the Inelastic scattering $\boldsymbol{e^{-}e^{+}}\rightarrow\boldsymbol{\mu^{-}\mu^{+}}$}\label{inelastic}
\begin{figure}[H]
\captionsetup{justification=centering}
\centering
\includegraphics[scale=0.47]{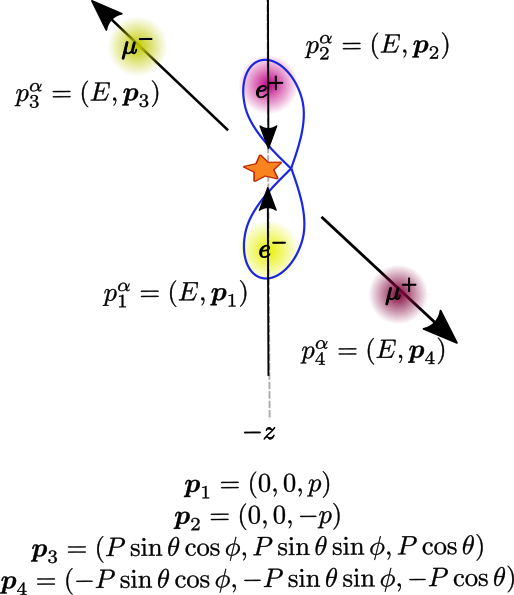} 
\caption{Collision diagram for the inelastic scattering $e^{-}e^{+}\rightarrow \mu^{-}\mu^{+}$.}
\label{fig:electro-muon}
\end{figure}

\begin{figure}[H]
\captionsetup{justification=centering}
\centering
\includegraphics[scale=0.35]{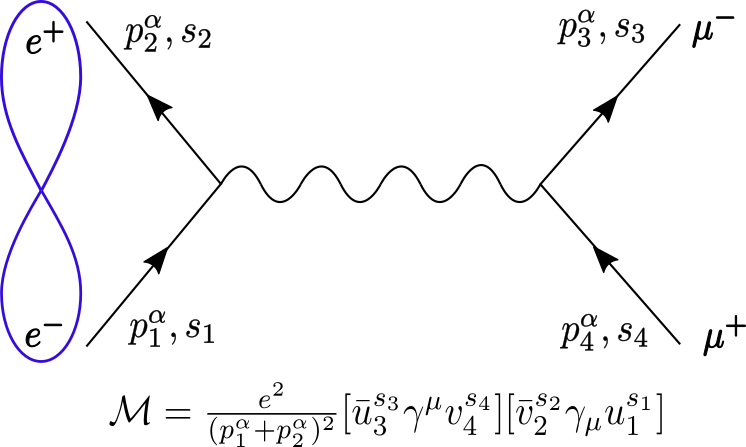} 
\caption{Feynman diagram for the inelastic scattering $e^{-}e^{+}\rightarrow \mu^{-}\mu^{+}$.}
\label{fig:feynmaninelastico}
\end{figure}

\begin{figure}[H]
\captionsetup{justification=centering}
\centering
\includegraphics[scale=0.25]{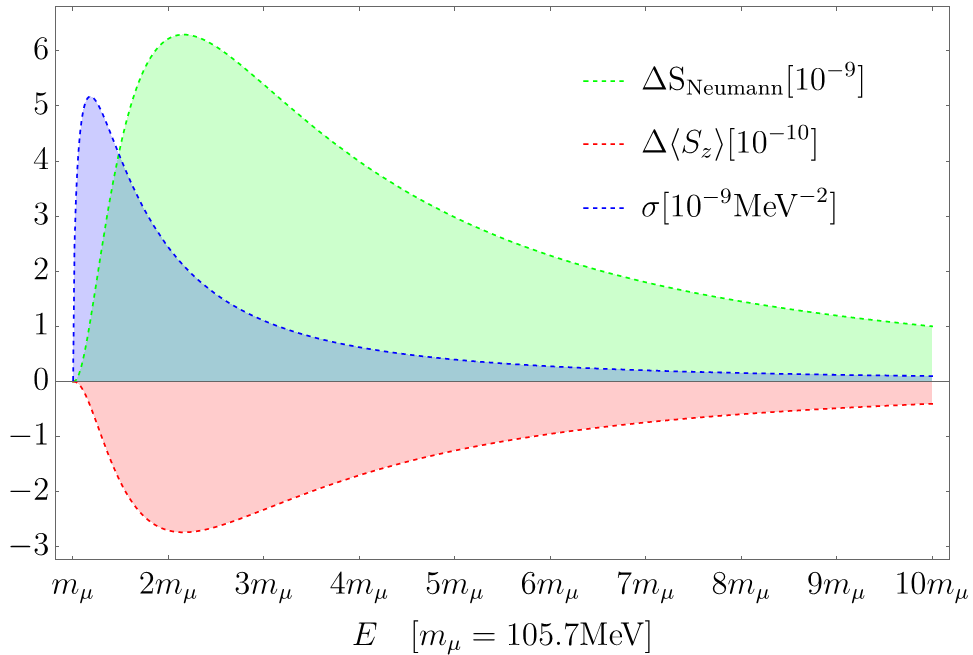} 
\caption{Variation of the von Neumann entropy and spin expectation value in the $z$ direction for $C$ as functions of the collision energy.}
\label{inelasticol}
\end{figure}

In the previous case, we carried out a non-relativistic expansion of the scattering amplitudes in the center-of-mass frame, keeping terms up to a specified order in the particle momenta, as dictated by to Eqs. \eqref{gamazero} and \eqref{gamai}. This was possible because those expressions characterized the spinors of one particle. In this case, however, we analyze the scattering of a particle-antiparticle pair $e^{-}e^{+}\rightarrow\mu^{-}\mu^{+}$ illustrated in Figs. \ref{fig:electro-muon} and \ref{fig:feynmaninelastico}. Once again, the dynamic variables $\theta$ and $\phi$ are the same as in the case of M\o ller scattering, with the exception that spinors $u_{1(3)}$ and $v_{2(4)}$ are associated with the particle and antiparticle, respectively.
In contrast to Møller scattering, for the production $\mu^{-}\mu^{+}$ pair to occur, there must be a minimum collision energy, which in this case corresponds to the muon mass $E=m_{\mu}=105.7$ MeV. The reason lies in the fact that the process $e^{-}e^{+}\rightarrow\mu^{-}\mu^{+}$ is inherently relativistic by nature. Accordingly, we retain all interaction contributions without performing a non-relativistic expansion, avoiding the use of spinor expressions expanded to a specific order in momentum; that is, all orders involved in the scattering amplitudes are considered. When evaluating an initially separable state of the $e^{-}e^{+}$ pair, we obtain a trend similar to that observed in the $e^{-}e^{-}$ process, as shown in Fig. \ref{inelasticol}. In this case, we include the total cross section, since it is integrable over its entire domain without presenting divergences in the involved kinematic variables. In Ref. \cite{Fonseca}, a direct proportionality was obtained between this quantity and the variations of entropy and spin expectation value, taking advantage of the fact that all these quantities reached their maximum at the same energy point ($E=1.18m_{\mu}$); however, this is not possible for this case because that point differs for each quantity.
On the other hand, when considering an initially entangled state between the pair $e^{-}e^{+}$, the variation of von Neumann entropy is 
\begin{equation}
\Delta\text{S}_{\text{Neumann}} = (\text{S}_{\text{Neumann}})_{\text{out}} - (\text{S}_{\text{Neumann}})_{\text{in}}=0.
\end{equation}
That is, correlations are developed as long as the particles start in a separable state; for an initial configuration in which the particles begin entangled, it is not possible to develop correlations given its maximum initial mixedness. This result is not compatible with those of \cite{jimbo}, where, given an initial entanglement after substitution $\eta\rightarrow\pi/4$, a variation of the von Neumann entropy yields  $\Delta\text{S}_{\text{Neumann}}\neq0$. Furthermore, for an initial separable state after substitution $\eta\rightarrow0$, the von Neumann entropy yields $\Delta\text{S}_{\text{Neumann}}\rightarrow\infty$ as $E\rightarrow m_{\mu}$. Such a result would be inconsistent with the expectation that, in the low-energy regime, the same qualitative behavior found at high energies -namely $\Delta\text{S}_{\text{Neumann}}\rightarrow0$- should be recovered. We repeated the calculation and obtained a result consistent with the behavior described above for both low and high energies, as shown in Fig. \ref{inelasticol}.

\section{Conclusions and perspectives}\label{conclusions}

It was found that, at tree-level order, the current-dipole and dipole-dipole interactions are responsible for creating correlations between two interacting particles that initially began separate. This also happened when including a witness particle $C$ in an initially entangled system of three particles, W-state, where the reduced density matrix of one particle before the scattering is not completely mixed. The preceding discussion was observed in the process(es) $e^{-}e^{-}(C)\rightarrow e^{-}e^{-}(C)$ by isolating each interaction through the approximation of the scattering amplitudes to zeroth- and first-order in the particle momenta, and in the  already relativistic process $e^{-}e^{+}\rightarrow\mu^{-}\mu^{+}$ where all orders were considered. This was evidenced through variations in the single-particle von Neumann entropies, also implying a variation in the spin expectation value along the $z-$axis in the absence of quantum coherence creation.  This is explicitly exemplified in the last case, where the behavior of the von Neumann entropy is consistent with the expected trend $\Delta\text{S}_{\text{Neumann}}\rightarrow0$ in the energy limits $E\rightarrow m_\mu$ and $E\rightarrow\infty$, as illustrated in Fig. \ref{inelasticol}. \\

For the case $e^{-}e^{-}\rightarrow e^{-}e^{-}$, initially described by a spin-separable state, a boost perpendicular to the collision axis between the particles was applied, naturally giving rise to a Wigner rotation in their spins. This implied the emergence of quantum coherence in the single density matrix, which could be measured through the appearance of a spin expectation value along the $x-$axis. 
The cause lies in the initial state measured by an observer in the moving frame rather than being a result inherent to the scattering process. From this, it follows that the correlations created by this process remained unchanged in both frames, with only a redistribution among the elements of the density matrix. Therefore, the von Neumann entropy was confirmed to be a Lorentz invariant. As an interesting result, the absence of off-diagonal terms generated by the scattering process was found to originate from $\mathcal{M}=0$, whereas for the Wigner rotations it arose from the cancellation of opposite contributions even when $\mathcal{M}\neq0$.\\

The correlations were found at tree level, which implies that the unitarity of the scattering operator was satisfied at this order of perturbation theory. Considering other orders, the production of virtual and real particles, like soft photons, is necessary for the scattering process to meet all the physical/mathematical requirements to be unitary at all orders. In that case, one might expect a possible contribution (either positive or negative) to the correlations developed in interacting/witness particles, along with the emergence of entanglement among soft photons that can be measured through the amount of quantum information.  
\begin{acknowledgments}
MS thanks CNPq for a research grant 304230/2023-2. IGP thanks for grant 306528/2023-1 from CNPq. BH acknowledges support from Fundação para a Ciência e Tecnologia (FCT) by projects 10.54499/UIDB/04564/2020 (\url{https://sciproj.ptcris.pt/157582UID}), and by 10.54499/UIDP/04564/2020 (\url{https://sciproj.ptcris.pt/157889UID}). JDF was financed in part by the Coordenação de Aperfeiçoamento de Pessoal de Nível Superior - Brasil (CAPES).
\end{acknowledgments}
\appendix
\section{Scattering Amplitudes and factors $\Lambda_{\text{AC}}^{i,j}(s)$}
\begin{align*}
    \mathcal{M}^{Co}(\uparrow\uparrow\rightarrow\uparrow\uparrow)=\mathcal{M}^{Co}(\downarrow\downarrow\rightarrow\downarrow\downarrow)=\frac{e^2 \cot\theta\csc\theta}{E^2-m^2},
\end{align*}
\begin{align*}
\mathcal{M}^{Co}&(\uparrow\downarrow\rightarrow\uparrow\downarrow)=\mathcal{M}^{Co}(\downarrow\uparrow\rightarrow\downarrow\uparrow)=\frac{e^2 (1-\cos\theta)^{-1}}{2\left(E^2-m^2\right)},
\end{align*}
\begin{align}
\mathcal{M}&^{Co}(\uparrow\downarrow\rightarrow\downarrow\uparrow)=\mathcal{M}^{Co}(\downarrow\uparrow\rightarrow\uparrow\downarrow)=-\frac{e^2(1+\cos\theta)^{-1}}{2\left(E^2-m^2\right)};\label{Coulomb}
\end{align}
\begin{align*}
\mathcal{M}^{CC}(\uparrow\uparrow\rightarrow\uparrow\uparrow)=\mathcal{M}^{CC}(\downarrow\downarrow\rightarrow\downarrow\downarrow)=\frac{e^2 \cot\theta\csc\theta}{m^2 },
\end{align*}
\begin{align*}
\mathcal{M}^{CC}(\uparrow\downarrow\rightarrow\uparrow\downarrow)=\mathcal{M}^{CC}(\downarrow\uparrow\rightarrow\downarrow\uparrow)=\frac{e^2 \cot ^2\left(\frac{\theta }{2}\right)}{4m^2},
\end{align*}
\begin{align}
\mathcal{M}^{CC}(\uparrow\downarrow\rightarrow\downarrow\uparrow)=\mathcal{M}^{CC}(\downarrow\uparrow\rightarrow\uparrow\downarrow)=-\frac{e^2 \tan ^2\left(\frac{\theta }{2}\right)}{4m^2};\label{CC}
\end{align}
\begin{align*}
    \mathcal{M}^{DD}(\uparrow\uparrow\rightarrow\uparrow\uparrow)=\mathcal{M}^{DD}(\downarrow\downarrow\rightarrow\downarrow\downarrow)=-\frac{e^2 \cos\theta}{4 m^2},
\end{align*}
\begin{align*}
    \mathcal{M}^{DD}(\uparrow\uparrow\rightarrow\downarrow\downarrow)=\frac{e^2 e^{2 i \phi } \cos\theta}{4 m^2},
\end{align*}    

\begin{align*}
    \mathcal{M}^{DD}(\downarrow\downarrow\rightarrow\uparrow\uparrow)=\frac{e^2 e^{-2 i \phi } \cos\theta}{4 m^2},
\end{align*}
\begin{align*}
    \mathcal{M}^{DD}(\uparrow\uparrow\rightarrow\uparrow\downarrow)= \mathcal{M}^{DD}(\uparrow\uparrow\rightarrow\downarrow\uparrow)=-\frac{e^2 e^{i \phi } \sin\theta}{4 m^2},
\end{align*}
\begin{align*}
    \mathcal{M}^{DD}(\uparrow\downarrow\rightarrow\uparrow\uparrow)=\mathcal{M}^{DD}(\downarrow\uparrow\rightarrow\uparrow\uparrow)=-\frac{e^2 e^{-i \phi } \sin\theta}{4 m^2},
\end{align*}
\begin{align*}
    \mathcal{M}^{DD}(\uparrow\downarrow\rightarrow\uparrow\downarrow)=\mathcal{M}^{DD}(\downarrow\uparrow\rightarrow\downarrow\uparrow)=\frac{e^2 (\cos\theta+2)}{4 m^2},
\end{align*}
\begin{align*}
    \mathcal{M}^{DD}(\uparrow\downarrow\rightarrow\downarrow\uparrow)= \mathcal{M}^{DD}(\downarrow\uparrow\rightarrow\uparrow\downarrow)=\frac{e^2 (\cos\theta-2)}{4 m^2},
\end{align*}
\begin{align*}
    \mathcal{M}^{DD}(\uparrow\downarrow\rightarrow\downarrow\downarrow)=\mathcal{M}^{DD}(\downarrow\uparrow\rightarrow\downarrow\downarrow)=\frac{e^2 e^{i \phi } \sin (\theta )}{4 m^2},
\end{align*}
\begin{align}
    \mathcal{M}^{DD}(\downarrow\downarrow\rightarrow\uparrow\downarrow)=\mathcal{M}^{DD}(\downarrow\downarrow\rightarrow\downarrow\uparrow)=\frac{e^2 e^{-i \phi } \sin (\theta )}{4 m^2};\label{DD}
\end{align}
\begin{align*}
    &\mathcal{M}^{CD}(\uparrow\uparrow\rightarrow\uparrow\downarrow)= \mathcal{M}^{CD}(\uparrow\uparrow\rightarrow\downarrow\uparrow)=\nonumber\\
    \mathcal{M}^{CD}(&\uparrow\downarrow\rightarrow\downarrow\downarrow)= \mathcal{M}^{CD}(\downarrow\uparrow\rightarrow\downarrow\downarrow)=\frac{e^2 e^{i \phi } \csc\theta}{2 m^2},
\end{align*}
\begin{align}
    &\mathcal{M}^{CD}(\uparrow\downarrow\rightarrow\uparrow\uparrow)=\mathcal{M}^{CD}(\downarrow\uparrow\rightarrow\uparrow\uparrow)=\nonumber\\
    \mathcal{M}^{CD}(&\downarrow\downarrow\rightarrow\uparrow\downarrow)=\mathcal{M}^{CD}(\downarrow\downarrow\rightarrow\downarrow\uparrow)=-\frac{e^2 e^{-i \phi } \csc\theta}{2 m^2}.\label{CD}
\end{align}
\begin{eqnarray}
	\Lambda^{1,1}_{\text{AC}}(s)&=&c_{1}^{2}|\mathcal{M}(\uparrow\uparrow\rightarrow\uparrow s)|^{2}+c_{3}^{2}|\mathcal{M}(\uparrow\downarrow\rightarrow\uparrow s)|^{2}\nonumber\\
	&+& c_{5}^{2}|\mathcal{M}(\downarrow\uparrow\rightarrow\uparrow s)|^{2}+
	c_{7}^{2}|\mathcal{M}(\downarrow\downarrow\rightarrow\uparrow s)|^{2}\nonumber\\
	&+&
	2c_{3}c_{5}\mathcal{M}(\uparrow\downarrow\rightarrow\uparrow s)\mathcal{\bar{M}}(\downarrow\uparrow\rightarrow\uparrow s),\nonumber\\
	\Lambda^{2,2}_{\text{AC}}(s)&=&c_{2}^{2}|\mathcal{M}(\uparrow\uparrow\rightarrow\uparrow s)|^{2}+c_{4}^{2}|\mathcal{M}(\uparrow\downarrow\rightarrow\uparrow s)|^{2} \nonumber \\ &+&c_{6}^{2}|\mathcal{M}(\downarrow\uparrow\rightarrow\uparrow s)|^{2}+c_{8}^{2}|\mathcal{M}(\downarrow\downarrow\rightarrow\uparrow s)|^{2} \nonumber\\
	&+&2c_{4}c_{6}\mathcal{M}(\uparrow\downarrow\rightarrow\uparrow s)\mathcal{\bar{M}}(\downarrow\uparrow\rightarrow\uparrow s)\nonumber,\\
	\Lambda^{3,3}_{\text{AC}}(s)&=&c_{1}^{2}|\mathcal{M}(\uparrow\uparrow\rightarrow\downarrow s)|^{2}+c_{3}^{2}|\mathcal{M}(\uparrow\downarrow\rightarrow\downarrow s)|^{2} \nonumber\\ &+&c_{5}^{2}|\mathcal{M}(\downarrow\uparrow\rightarrow\downarrow s)|^{2}+c_{7}^{2}|\mathcal{M}(\downarrow\downarrow\rightarrow\downarrow s)|^{2}\nonumber\\
	&+&2c_{3}c_{5}\mathcal{M}(\uparrow\downarrow\rightarrow\downarrow s)\mathcal{\bar{M}}(\downarrow\uparrow\rightarrow\downarrow s),\nonumber\\
	\Lambda^{4,4}_{\text{AC}}(s)&=&c_{2}^{2}|\mathcal{M}(\uparrow\uparrow\rightarrow\downarrow s)|^{2}+c_{4}^{2}|\mathcal{M}(\uparrow\downarrow\rightarrow\downarrow s)|^{2} \nonumber\\ &+&c_{6}^{2}|\mathcal{M}(\downarrow\uparrow\rightarrow\downarrow s)|^{2}+c_{8}^{2}|\mathcal{M}(\downarrow\downarrow\rightarrow\downarrow s)|^{2}\nonumber\\
	&+&2c_{4}c_{6}\mathcal{M}(\uparrow\downarrow\rightarrow\downarrow s)\mathcal{\bar{M}}(\downarrow\uparrow\rightarrow\downarrow s),\nonumber\\
	\Lambda^{1,2}_{\text{AC}}(s)&=&c_{1}c_{2}|\mathcal{M}(\uparrow\uparrow\rightarrow\uparrow s)|^{2}+c_{3}c_{4}|\mathcal{M}(\uparrow\downarrow\rightarrow\uparrow s)|^{2} \nonumber \\ &+&c_{5}c_{6}|\mathcal{M}(\downarrow\uparrow\rightarrow\uparrow s)|^{2}+c_{7}c_{8}|\mathcal{M}(\downarrow\downarrow\rightarrow\uparrow s)|^{2}
	\nonumber\\
	&+&(c_{3}c_{6}+c_{4}c_{5})\mathcal{M}(\uparrow\downarrow\rightarrow\uparrow s)\mathcal{\bar{M}}(\downarrow\uparrow\rightarrow\uparrow s),\nonumber\\
	\Lambda^{3,4}_{\text{AC}}(s)&=&c_{1}c_{2}|\mathcal{M}(\uparrow\uparrow\rightarrow\downarrow s)|^{2}+c_{3}c_{4}|\mathcal{M}(\uparrow\downarrow\rightarrow\downarrow s)|^{2}\nonumber\\
	&+&c_{5}c_{6}|\mathcal{M}(\downarrow\uparrow\rightarrow\downarrow s)|^{2}+c_{7}c_{8}|\mathcal{M}(\downarrow\downarrow\rightarrow\downarrow s)|^{2}\nonumber\\
	&+&(c_{3}c_{6}+c_{4}c_{5})\mathcal{M}(\uparrow\downarrow\rightarrow\downarrow s)\mathcal{\bar{M}}(\downarrow\uparrow\rightarrow\downarrow s).\label{rhosA}
\end{eqnarray}

\end{document}